# Android Malware Category and Family Detection and Identification using Machine Learning

Ahmed Hashem El Fiky[1*], Ayman El Shenawy[1, 2], Mohamed Ashraf Madkour[1]
1 Systems and Computer Engineering Dept., Faculty of Engineering, Al-Azhar University, Cairo, Egypt.

2 Software Engineering and Information Technology, Faculty of Engineering and Technology, Egyptian Chinese University, Cairo, Egypt.
0x4186@gmail.com Eaymanelshenawy@azhar.edu.eg mamadkour@gmail.com

## Abstract:

Android malware is one of the most dangerous threats on the internet, and it's been on the rise for several years. Despite significant efforts in detecting and classifying android malware from innocuous android applications, there is still a long way to go. As a result, there is a need to provide a basic understanding of the behavior displayed by the most common Android malware categories and families. Each Android malware family and category has a distinct objective. As a result, it has impacted every corporate area, including healthcare, banking, transportation, government, and e-commerce. In this paper, we presented two machine-learning approaches for Dynamic Analysis of Android Malware: one for detecting and identifying Android Malware Categories and the other for detecting and identifying Android Malware Families, which was accomplished by analyzing a massive malware dataset with 14 prominent malware categories and 180 prominent malware families of CCCS-CIC-AndMal2020 dataset on Dynamic Layers. Our approach achieves in Android Malware Category detection more than 96 % accurate and achieves in Android Malware Family detection more than 99% accurate. Our approach provides a method for high-accuracy Dynamic Analysis of Android Malware while also shortening the time required to analyze smartphone malware.

## Introduction:

Android dominates the smartphone and operating system markets, with market shares at all levels. It grew from a minor player when it debuted in 2010, to powering 87 percent of smartphones globally in 2019, and it appears that Android's dominance will continue in the coming years, with its smartphone share expected to increase to 87.4 percent in 2023 [1]. Furthermore, it controls more than 41.42 percent of the global operating system market [2]. At the same time, the Android malware industry is being increased, with nearly 12,000 new cases of Android malware being reported every day [3]. The current evolutionary process will soon result in malware that uses adaptive and success-based learning to improve attack effectiveness. Unquestionably, better malware detection tools are required.

Android malware is malicious software that designed to target smartphone devices that run the Android operating system. It behaves similarly to other malware samples that run on desktop or laptops. Android malware is also known as mobile malware, and it is any malicious software that designed to harm the mobile device by performing illegal activities. Adware [4], backdoor, file infector, potentially unwanted application (PUA), ransomware [5], riskware, scareware [6], and Trojan [7] are some of the malware categories. Each malware category has distinguishing features that set it apart from the others. Android malware, like humans, evolves. There are Malware families are associated with each malware category.

The unrivalled threat of Android malware is the root cause of a slew of internet security issues, posing a new challenge to researchers and cybersecurity experts. The only way to eliminate this threat is to detect and mitigate malware samples as soon as they detected. The key to accomplishing this is fundamental knowledge of Android malware categories and families. The purpose of this paper is to shed light on prominent Android malware categories as well as related families within each malware category. Furthermore, it familiarizes the reader with the abnormal activities carried out by each malware category. Finally, the paper presents two machine-learning models for detecting and classifying Android malware categories and families. The following contributions made in this paper:

1. Propose an effective systematic and functional approach to detect and identify Android malware category, and family on Dynamic layers to avoid the limitations of Static Analysis, such as the inability to detect the Android Malware app that used obfuscation techniques.
2. Evaluate the effectiveness of our model using machine-learning classifiers.
3. Present Comparison between machine learning techniques: Decision tree (J48), Naive Bayesian (NB), Support Vector Machine (SVM), AdaBoost (AB), Logistic Regression (LR), K-Nearest Neighbor (KNN), Random Forest (RF), and Multilayer Perceptron (MLP).
4. Present Comparison between our work and that of others.

The remainder of the paper structured as follows. Section 2 provides a synopsis of the related work on Android malware detection. Section 3 contains a detailed discussion of the chosen dataset. Section 4 provides an overview of our proposed model for detecting and identifying Android malware categories and families. Section 5 contains the experimental setup and results. Section 6 contains a comparative analysis. Section 7 concludes with recommendations for future work.

## Related Work

The emergence of sophisticated open source tools for disseminating persistent zero-day malware that evades security measures has paved the way for various types of malware analysis. Furthermore, security breaches in the Android framework mostly caused by the installation of third-party applications. More advanced methods for static and dynamic analyses have recently been proposed, attempting to incorporate the benefits of traditional methods while improving on their limitations. This section highlights some of the most recent work on android malware detection and classification.

The authors of [8] built three distinct types of datasets based on Machine Learning using a control flow graph and API information taken from Android Malware applications. To achieve 98.98 % detection precision, DNN, LSTM, and C4.5 algorithms used to run tests on 10,010 benign and 10,683 malicious applications. The authors of [9] employed static and code analysis approaches to identify the spare permissions sought by some apps to do suspicious actions, and they were able to detect Android Malicious apps with 91.95 % accuracy.

The accuracy of identifying ransomware with machine learning techniques was examined by the authors in [10] using the CICAndMal2017 dataset of 10 ransomware families. The CICAndMal2017 dataset [11] contains both benign and malicious applications. Adware, Ransomware, Scareware, and SMS Malware are the four types of malware found in CICAndMal2017. There are over 80 network traffic features in this dataset. The researchers used the CICAndMal2017 dataset in [12], and they chose one PCAP file for each malware family at random. Then, in two steps, they extracted features from PCAP files. The first step was to isolate network flows using a Java software that used a flow-level method. After that, 15 features retrieved and three supervised machine-learning classifiers utilized. Applications classified as malware, benign, adware, ransomware, or scareware using Random Forest, K-Nearest Neighbors, and Decision Tree.

The CICIDS2017 dataset was created using data collected between 3 and 7 July 2017 [13]. The authors proposed the CICID2017 dataset in [14] because it worked for web, Denial of Service (Dos), Distributed Denial of Service (DDoS), Infiltration, Brute Force Secure Shell (SSH), and File Transfer Protocol (FTP) assaults. CICFlowMeter examined the generated network traffic and collected 80 network flow features. The CICIDS2017 dataset retrieved the basic behavior of twenty-five users using FTP and Hypertext Transfer Protocol (HTTPS). Using the CICAndMal2017 dataset, the authors in [15] assessed the performance measures for several detectors.

TFDroid [17] utilized SVM to detect malware in android apps by looking at the source, sink, and description. Outliers identified by clustering applications into distinct domains depending on the description. Cross-validation performed on a small dataset, and only benign applications utilized to train the classifier, resulting in a 93.65% accuracy in recognizing the malicious program. To detect unfamiliar APKs, Suman et al. [18] used a combination of sensitive permissions and API attributes, as well as an ensemble-learning model based on decision tree and KNN classifiers. Zhang et al. [19] created behavioral semantics by calculating the confidence of association rules between abstracted API calls to define an application and then detecting them using different machine learning techniques (KNN, RF, SVM). The model outperformed MaMaDroid [20], which created with the same attributes and machine learning algorithms as the model.

Suleiman et al. [21] used static features (permissions, intents, API calls, and date of appearance) derived from date-labeled benign and malware datasets to evaluate the effectiveness of various machine learning classifiers (NB, J48, SVM, RF, SL). GefDroid [22] used unsupervised learning to do graph embedding based on android malware family analysis. The program semantics abstracted using a fine-grained behavioral model to create a set of sub-graphs. Intentions and permissions employed by AndroDialysis [23] to detect malware. It compared the efficacy of using permissions with intent. The Bayesian Network algorithm created in order to attain a greater detection rate and faster detection time.

To characterize the app's activities, Wang et al. [24] retrieved 11 types of static characteristics from each app, primarily from API calls, permissions, intents, and hardware information. To detect malware and categorize innocent apps, it used an ensemble of various classifiers including SVM, KNN, NB, CART, and RF. For static Android malware analysis, Atici et al. [25] suggested an approach based on control flow graphs and machine learning algorithms (CART, Probabilistic Neural Network (PNN), NB, and 1-NN).

After training with a set of benign and malicious apps, ICCDetector [26] uses a detection model to classify Android malware data into five established malware categories. It uses a trained malware detection model that tested using SVM on 5,264 malicious and 12,026 benign apps. A machine learning-based Android malware detection and family identification solution proposed by Garcia et al. [27]. Its features based on categorized Android API usage, reflection-based features, and features from native app binaries. Based on a bi-objective GAN, Li et al. [28] created a new adversarial example attack approach. Permissions, actions, and all API calls employed as features.

## Dataset and Preliminaries

We used the recently released Android Malware Dataset, CCCS-CIC-AndMal-2020, [30], which comprises 400K android apps, to test and assess our suggested methodology (200K benign and 200K malware). Android Malware apps evaluated in a virtual environment. This dataset includes static elements such as Android malware families, permissions, and intents, as well as dynamic features like as API calls and all generated log files. Process logs, packages, log states, battery states, and other collected information are also included in the dataset. Table 2 provides a full description of the datasets adapted from [31] that are locally available and include published year data. Table 3 compares the properties of our dataset to those of other freely available datasets. Adware, backdoors, file infectors, no-category, Potentially Unwanted Apps (PUA), ransomware, riskware, scareware, Trojan, trojan-banker, trojan-dropper, trojan-sms, trojan-spy, and zero-day are the 14 Category of malware found in this dataset. We concentrated on the dataset's Dynamic Analysis section, which is broken into two parts: the first is Dynamic Analysis of Android Malware before rebooting an Android emulator, with 28,380 Android Malware apps as shown in Table 4. The second section is Dynamic Analysis of Android Malware after reboot Android emulator, including 25,059 Android Malware apps. Because the number of applications in first part bigger than the second portion, we used the first part (before reboot) in our test.

This section goes over the functions that each of these malware categories does, as well as some of the most important malware families that fall under each of these malware categories [32]. We have illustrated the main Android Malware groups using graphics in Figure 1 and Table 1.

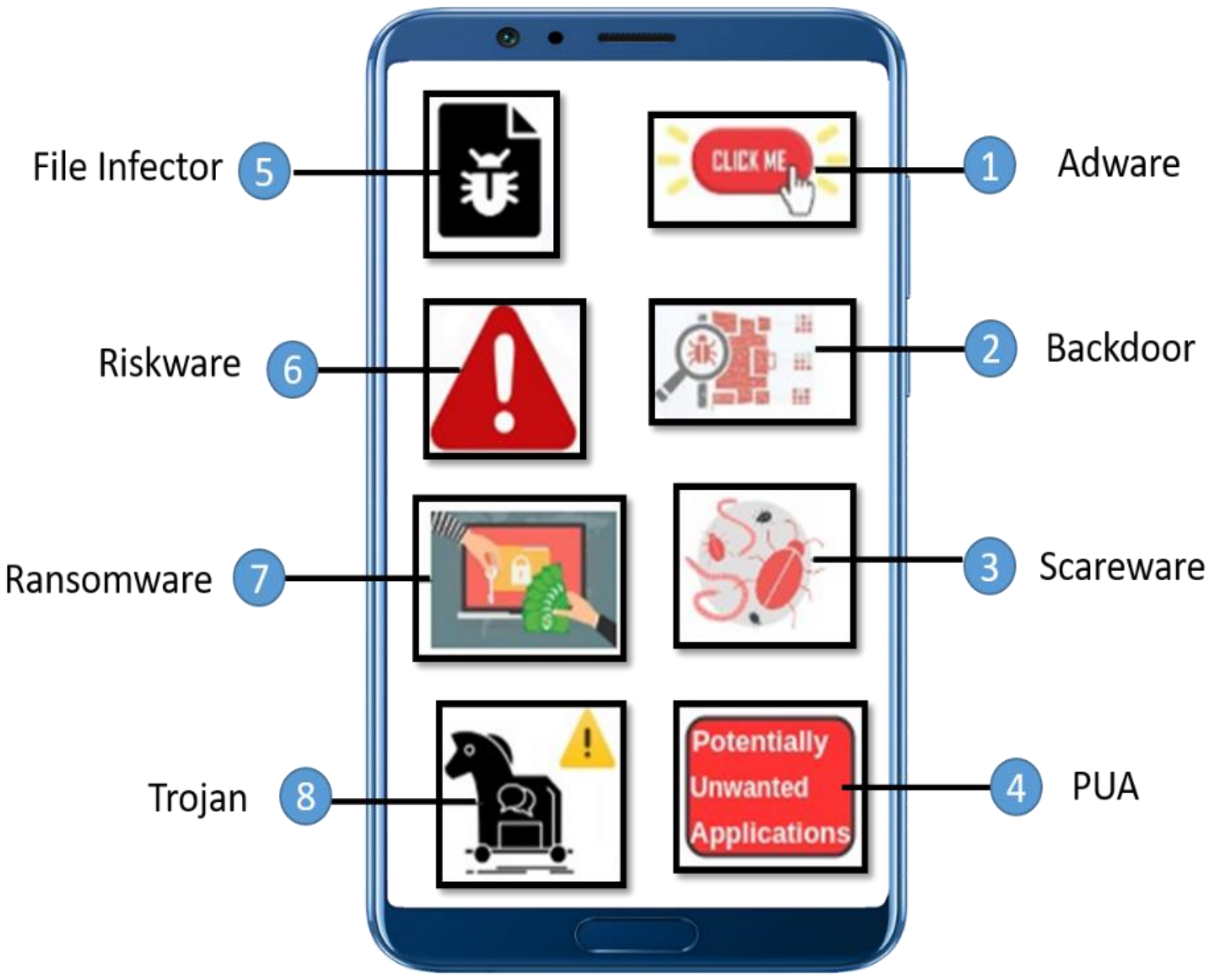

Figure 1: Main Android Malware Categories

Table 1: Main Android Malware Categories Activities

| Malware Category | Malware Definition | Malware Activities |
|---|---|---|
| Adware | It's a malicious program that displays intrusive adverts on the user's screen, particularly when using web services | Adware gathers personal information from the device in general, including phone numbers, email addresses, application accounts, the device's IMIE number, device ID, and status. Some adware families use the camera on the smartphone to capture images. Adware can attempt to encrypt data on devices and install other harmful programs, code, or files. |
| Backdoor | Backdoors are unnoticed entrances into a smartphone. Backdoors, in other words, are a means to circumvent a smartphone's authentication and elevate privileges, enabling an attacker to access the device at any time | Backdoor malware captures personal information from the phone, sends/receives messages, makes calls and records call history, collects lists of installed and running programs, and frees up memory on the device. The backdoor is rooted to the Android device on which it was installed in some extreme situations. Adware and backdoors are sometimes related. Advertisement malware is frequently used by attackers to entice victims. A backdoor is placed on the user's device once they click the advertising. |
| Scareware | Scareware is a fear inducer that encourages users to download or purchase harmful applications by instilling dread in their brains. For example, persuading people to install a phony software that claims to protect the device. | Scareware try to acquire information about the device, including its GPS position, and then install malicious malware on it. |
| PUA | PUAs are potentially unwanted applications that packaged with legitimate, free-to-download software. They're also known as possibly unwanted programs (PUPs). PUAs are not necessarily harmful. It all depends on how they're going to be used. Adware, spyware, and hijackers are examples of this type of malware. | PUAs exploit the device to gather personal information and user contacts. They can utilize the Global Positioning System (GPS) to track the device's position and show pop-up advertising, notifications and warnings, undesirable URLs, and shortcuts on the user's screen. |
| File Infector | file infector is a type of malware that infects APK files. Android Package Kit (APK) is a file that contains all of an application's data. With the help of APK files, the file infector is set up. After that, when APK files are installed, the virus is run. | file infector tries to slow down the device and use a lot of battery power. Device ID, IMEI number, and phone status are all collected by these families. They have the ability to ban, remove, and utilize phone applications. They have access to files and device settings and may alter, gather, and access them. In the worst-case scenario, file infector can get root access to the device. |
| Riskware | Riskware is a genuine software that poses a threat to the device's security vulnerabilities. Despite the fact that it is a legitimate software, it is used to collect data from the device and lead users to malicious websites. | Riskware families acquire personal and phone information, send and receive SMSs, steal network information, link to malicious websites, install harmful content on devices, display malicious advertising, and change system settings and files. |
| Ransomware | Ransomware is a type of malware that encrypts files and folders on the machine to prevent users from accessing them. It demands a hefty ransom in exchange for the decryption key that would allow the data to be unlocked. | Sending and receiving SMSs, locking SIM cards and cellphones, collecting network information such as Wi-Fi connection data, and interacting with the remote server managing the ransomware assault are all part of ransomware families. |
| Trojan | Trojans are sneaky impersonators that masquerade as genuine programs. They can remain undetected in the background while stealing data from the device. | Trojans frequently delete, alter, block, and copy data in order to impair operating system services. |

Table 2: Details regarding currently available android malware datasets [31].

| Dataset | Pub. Year | No. of Benign | No. of Malware | Captured static features | | | | | | | Captured dynamic features | | | | | | Installed On |
|---|---|---|---|---|---|---|---|---|---|---|---|---|---|---|---|---|---|
| | | | | S | P | I | C | Ce | SC | MD | APC | Net | Sys.C | IF | Log | M | |
| [33] | 2012 | - | 1,260 | N | Y | N | Y | N | Y | N | N | N | N | N | N | N | - |
| [34] | 2014 | 123,453 | 5,560 | N | Y | Y | Y | N | Y | N | N | N | N | N | N | N | - |
| [35] | 2015 | 51,179 | 4,554 | N | Y | Y | N | Y | Y | N | N | N | N | N | N | N | - |
| [36] | 2016 | 1,776 | 906 | N | N | N | N | N | N | N | Y | N | N | N | N | N | EM |
| [37] | 2016 | 1,776 | 906 | N | Y | Y | N | Y | Y | N | Y | N | N | N | N | N | EM |
| [38] | 2016 | 8,840 | 643 | N | N | N | N | N | N | N | N | N | Y | N | Y | N | EM |
| [39] | 2016 | - | 7 | N | N | N | N | N | N | N | N | N | N | Y | N | N | RP |
| [40] | 2017 | 1,500 | 400 | N | N | N | N | N | N | N | N | Y | N | N | N | N | RP |
| [41] | 2017 | - | 405 | N | N | N | Y | N | Y | N | N | N | N | N | N | N | - |
| [42] | 2018 | 38,000 | 33,000 | N | N | N | N | N | Y | N | N | N | N | N | N | N | - |
| [43] | 2018 | 1,2M | - | N | N | N | Y | N | Y | N | N | N | N | N | N | N | - |
| [11] | 2018 | 1,700 | 426 | Y | Y | Y | N | N | N | N | Y | Y | N | N | N | N | RP |
| [45] | 2019 | 5,065 | 426 | Y | Y | Y | N | N | N | N | Y | Y | N | N | N | N | RP |
| [30] | 2020 | 200K | 200K | Y | Y | Y | Y | N | N | Y | Y | Y | Y | N | Y | Y | EM |

Legend:

S-States, P-Permissions, I-Intents, C-Components, Ce-Certificate, SC-Source Code, MD-Metadata, APC-API Call, Net-Network, Sys.C-System Call, IF-Infoflow, M-Memory, EM-Emulator, RP-Real Phone, N-No, Y-Yes

Table 3: Comparison of publicly available android malware datasets [31].

| Dataset | Pub. Year | A1 | A2 | A3 | A4 | A5 | A6 | A7 | A8 | A9 | A10 | A11 | A12 | A13 | A14 | A15 |
|---|---|---|---|---|---|---|---|---|---|---|---|---|---|---|---|---|
| [33] | 2012 | S | - | - | Y | - | - | Y | Y | - | Y | Y | N | Y | Y | N |
| [34] | 2014 | S | - | - | Y | - | - | Y | Y | N | Y | Y | Y | Y | N | N |
| [35] | 2015 | S | - | - | Y | - | - | N | Y | N | Y | Y | Y | Y | N | N |
| [36] | 2016 | B | N | Y | Y | N | N | Y | Y | Y | Y | N | N | Y | N | N |
| [37] | 2016 | B | N | Y | Y | N | N | Y | Y | Y | Y | N | Y | Y | N | N |
| [38] | 2016 | B | N | Y | Y | N | N | Y | N | Y | Y | Y | N | Y | N | N |
| [39] | 2016 | B | Y | N | Y | Y | N | Y | N | - | Y | N | N | Y | N | Y |
| [40] | 2017 | D | Y | Y | Y | Y | N | Y | N | Y | Y | Y | N | Y | N | Y |
| [41] | 2017 | S | - | - | Y | - | - | Y | Y | - | Y | Y | Y | Y | Y | Y |
| [42] | 2018 | S | N | N | Y | N | N | Y | Y | Y | N | Y | N | Y | N | Y |
| [43] | 2018 | S | N | N | Y | N | N | Y | Y | N | N | Y | N | N | N | Y |
| [11] | 2018 | B | Y | Y | Y | Y | Y | Y | Y | Y | Y | Y | Y | Y | Y | N |
| [45] | 2019 | B | Y | Y | Y | Y | Y | Y | Y | Y | Y | Y | Y | Y | Y | Y |
| [30] | 2020 | B | N | Y | Y | Y | Y | Y | Y | Y | Y | Y | Y | Y | Y | Y |

Legend:

A1: Type of data capturing, Static(S) or Dynamic(D) or both(B).

A2: Utilizing Real-Phone devices instead of emulators.

A3: Having network architecture for the experiment set up.

A4: Examining real-world malware samples.

A5: Having malware activation scenario.

A6: Defining multiple states of data capturing.

A7: Having trust-able fully-labeled malware samples.

A8: Including diverse malware categories and families.

A9: Keeping balance between malicious and benign samples.

A10: Avoiding anonymity and preserving all captured data.

A11: Containing a heterogeneous set of resources.

A12: Providing a variety of feature sets for other researchers.

A13: For meta-data, includes a proper documentation.

A14: Including malware taxonomy.

A15: Being up-to-date.

N-No, Y-Yes

Table 4: Dynamic Analysis (before reboot) Dataset details

| Malware Category | No. of Malware Family | No. of Samples | Common Malware Family |
|---|---|---|---|
| Adware | 43 | 5838 | gexin, batmobi, ewind, shedun, pandaad, appad, dianjin, gmobi, hummingbird, mobisec, loki, kyhub, and adcolony |
| Backdoor | 11 | 591 | mobby, kapuser, hiddad, dendroid, levida, fobus, moavt, androrat, kmin, pyls, and droidkungfu |
| File Infector | 5 | 129 | leech, tachi, commplat, gudex, and aqplay |
| No_Category | 1 | 1048 | No_Category |
| PUA | 9 | 665 | apptrack, secapk, wiyun, youmi, scamapp, utchi, cauly, and umpay |
| Ransomware | 8 | 1861 | congur, masnu, fusob, jisut, koler, lockscreen, slocker, and smsspy |
| Riskware | 19 | 7261 | badpac, mobilepay, wificrack, triada, skymobi, deng, jiagu, smspay, smsreg, and tordow |
| Scareware | 4 | 462 | avpass, mobwin, and fakeapp |
| Trojan | 38 | 4412 | gluper, lotoor, rootnik, guerrilla, gugi, hqwar, obtes, and hypay |
| Trojan_Banker | 11 | 118 | minimob, marcher, faketoken, zitmo, bankbot |
| Trojan_Dropper | 9 | 837 | Ramnit, cnzz, rooter, gorpo, xiny |
| Trojan_SMS | 10 | 1028 | Opfake, plankton, boxer, vietsms |
| Trojan_Spy | 11 | 1801 | Smsthief, qqspy, spyagent, smforw |
| Zero_Day | 1 | 2329 | Zero_Day |
| Sum | 180 | 28,380 | |

## Android Malware Category, Family detection and identification

This section outlines our suggested method for detecting malicious applications. Our method entails feature extraction and selection, as well as classification of Android Malware apps by Malware Category and Malware Family. Our suggested solution, which consists of Dynamic malware categorization for Android Malware category and family, is summarized in Figure 2. Where Android Malware applications are categorized into 14 categories (adware, backdoor, file infector, no-category, Potentially Unwanted Apps (PUA), ransomware, riskware, scareware, trojan, trojan-banker, trojan-dropper, trojan-sms, trojan-spy, and Zero-day) and 180 families simultaneously.

### Pre-Processing

Pre-processing is a procedure taken before a machine learning model is used to achieve the best results. NaN removal, duplicate instances removal, and normalization/scaling are all common pre-processing procedures. We chose MinMax scaling for feature normalization since the given dataset has minimal variance and ambiguity. The rescaling of real-valued numeric characteristics to a defined range is referred to as normalization (e.g., 0 and 1). When using a model that relies on the magnitude of values, scaling the input characteristics is critical. MixMax scaling normalizes data using the formula mentioned in (1).

$$X_{norm} = \frac{X_i - X_{min}}{X_{max} - X_{min}} \qquad (1)$$

where $X_i$ is the original value of the feature that is subtracted from minimum value of that feature and divided by subtracted result of maximum and minimum of the feature.

**Feature Extraction and Selection**

After executing the malware in an emulated environment, six types of characteristics extracted to better understand the behavioral changes of these malware categories and families. The following are the major characteristics that were extracted:

- Memory: Memory features define activities performed by malware by utilizing memory.
- API: Application Programming Interface (API) features delineate the communication between two applications.
- Network: Network features describe the data transmitted and received between other devices in the network. It indicates foreground and background network usage.
- Battery: Battery features describe the access to battery wakelock and services by malware.
- Logcat: Logcat features write log messages corresponding to a function performed by malware.
- Process: Process features count the interaction of malware with total number of process.

The number of extracted features is 141 features (Large number). So, we used feature selection methods like Chi2 and Mutual Information (MI) to select important features, remove redundancy and Irreverent features to enter in classification process whether for category classification or for family classification.

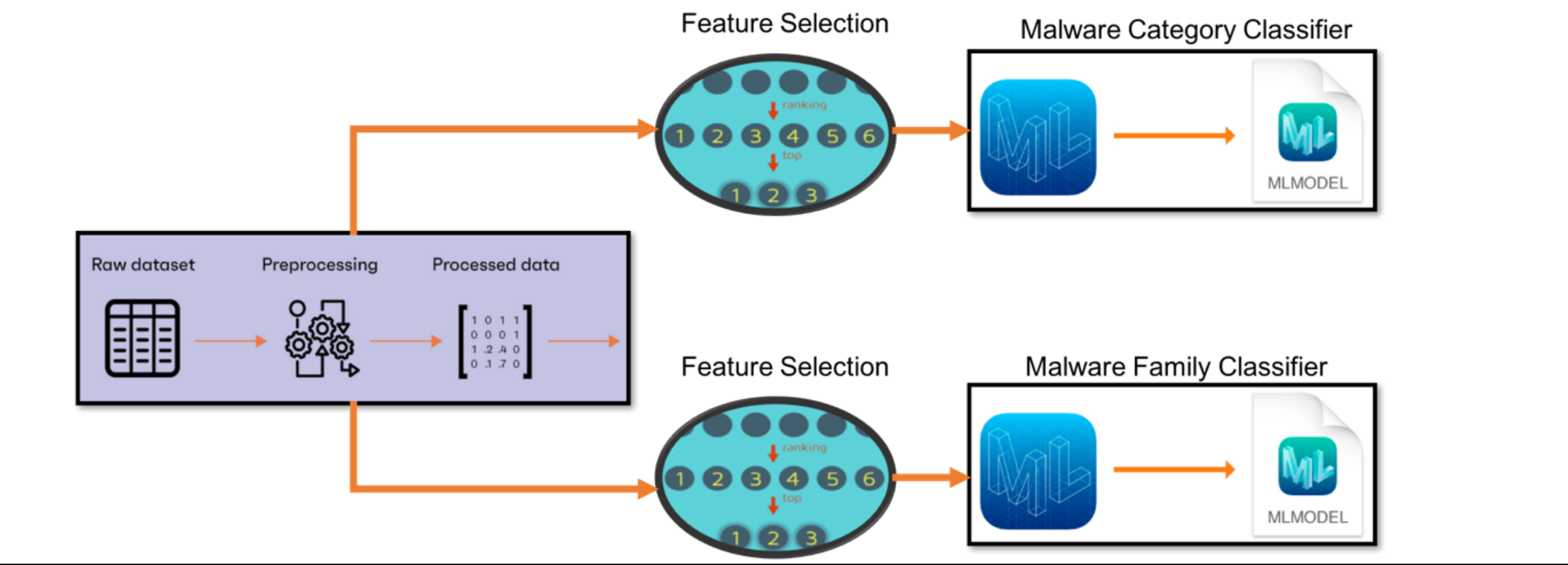

Figure 2: Graphical representation of proposed android malware application detection and identification approach.

**Machine Learning Models**

In this study, we use the following machine learning algorithms to evaluate and compare the effectiveness of our proposed approach: Decision tree (J48), Naive Bayesian (NB), Support Vector Machine (SVM), AdaBoost (AB), Logistic Regression (LR), K-Nearest Neighbor (KNN), Random Forest (RF), and Multilayer Perceptron (MLP).

## Evaluation and Results

Evaluation measure and experimental setup play a vital role to analyze the performance of a machine learning model. For the experiment, we used cross validation (K=10). We computed accuracy, recall, False Positive Ratio, and False Negative Ratio to evaluate the performance of the proposed method. Details can be seen in the Equations given below. Table 5 presents the computing environment used for experiments.

$$Accuracy = \frac{TP+TN}{TP+TN+FP+FN} \quad (2)$$

$$Recall = \frac{TP}{TP+FN} \quad (3)$$

$$FPR = \frac{FP}{FP+TN} \quad (4)$$

$$FNR = \frac{FN}{TP+FN} \quad (5)$$

Here

- True Positives (TP): Predicted case to be in YES and is actually in it.
- False Positives (FP): Predicted case to be in YES but is not actually in it.
- True Negative (TN): Predicted case not to be in YES and is not actually in it.
- False Negative (FN): Predicted case not to be in YES, but is actually in it.

Table 5: Computing Environment

| Parameter | Value |
|---|---|
| Operating System | MacOS High Sierra v 10.13.4 |
| CPU | 2.8 GHz Intel Core i7 |
| RAM | 16 GB |
| Python Version | 2.7 |

### Malware Category Detection

1- Comparison between Feature selection methods Chi2 and MI as shown in Figure 3, 4 and Table 6,7 to determine the best feature selection method and determine the best threshold.

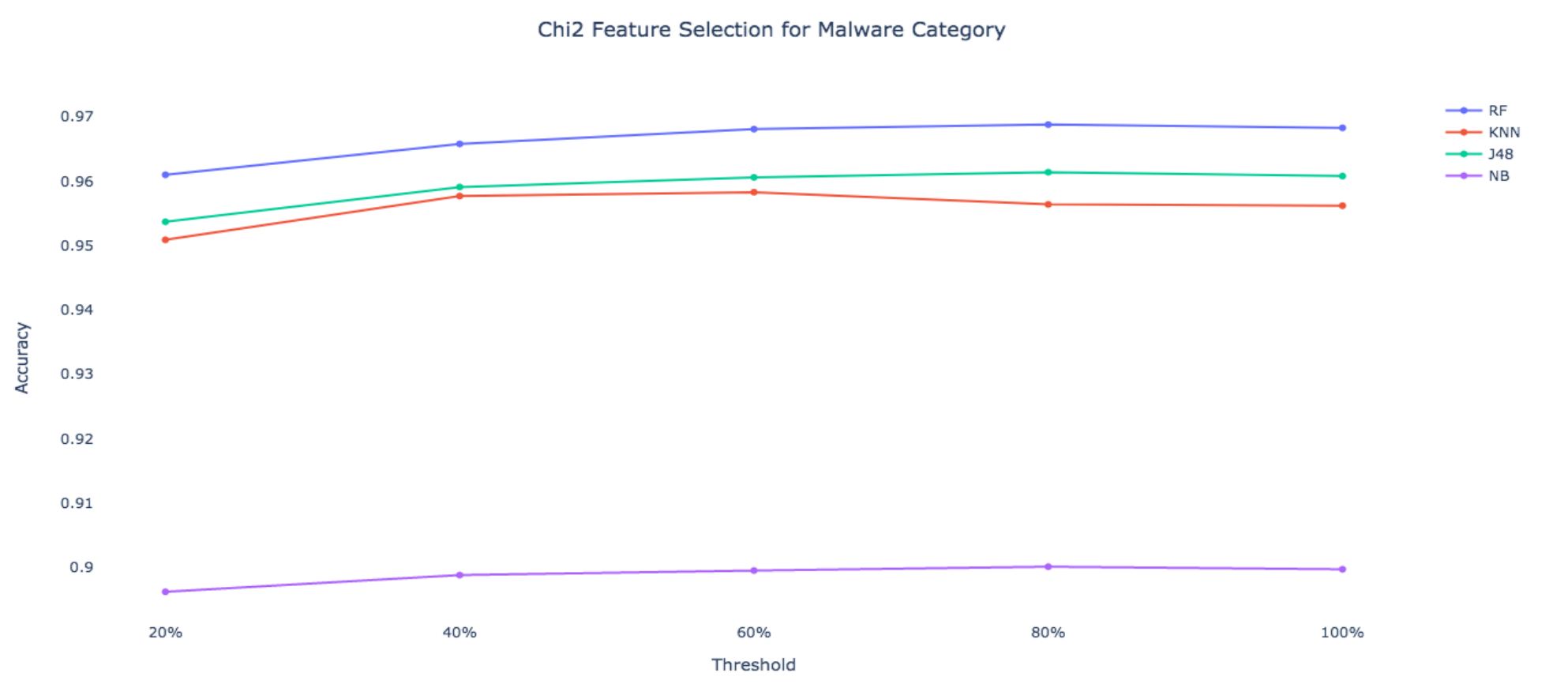

Figure 3: Chi2 Feature Selection for Malware Category

Table 6: Comparison between accuracy and Threshold percentage for Chi2 Feature Selection for Malware Category

| Threshold | Accuracy | | | |
|---|---|---|---|---|
| | RF | KNN | J48 | NB |
| 20% | 0.961 | 0.9509 | 0.9537 | 0.8962 |
| 40% | 0.9658 | 0.9577 | 0.9591 | 0.8988 |
| 60% | 0.9681 | 0.9583 | 0.9606 | 0.8995 |
| 80% | 0.9688 | 0.9564 | 0.9614 | 0.9001 |
| 100% | 0.9683 | 0.9562 | 0.9608 | 0.8997 |

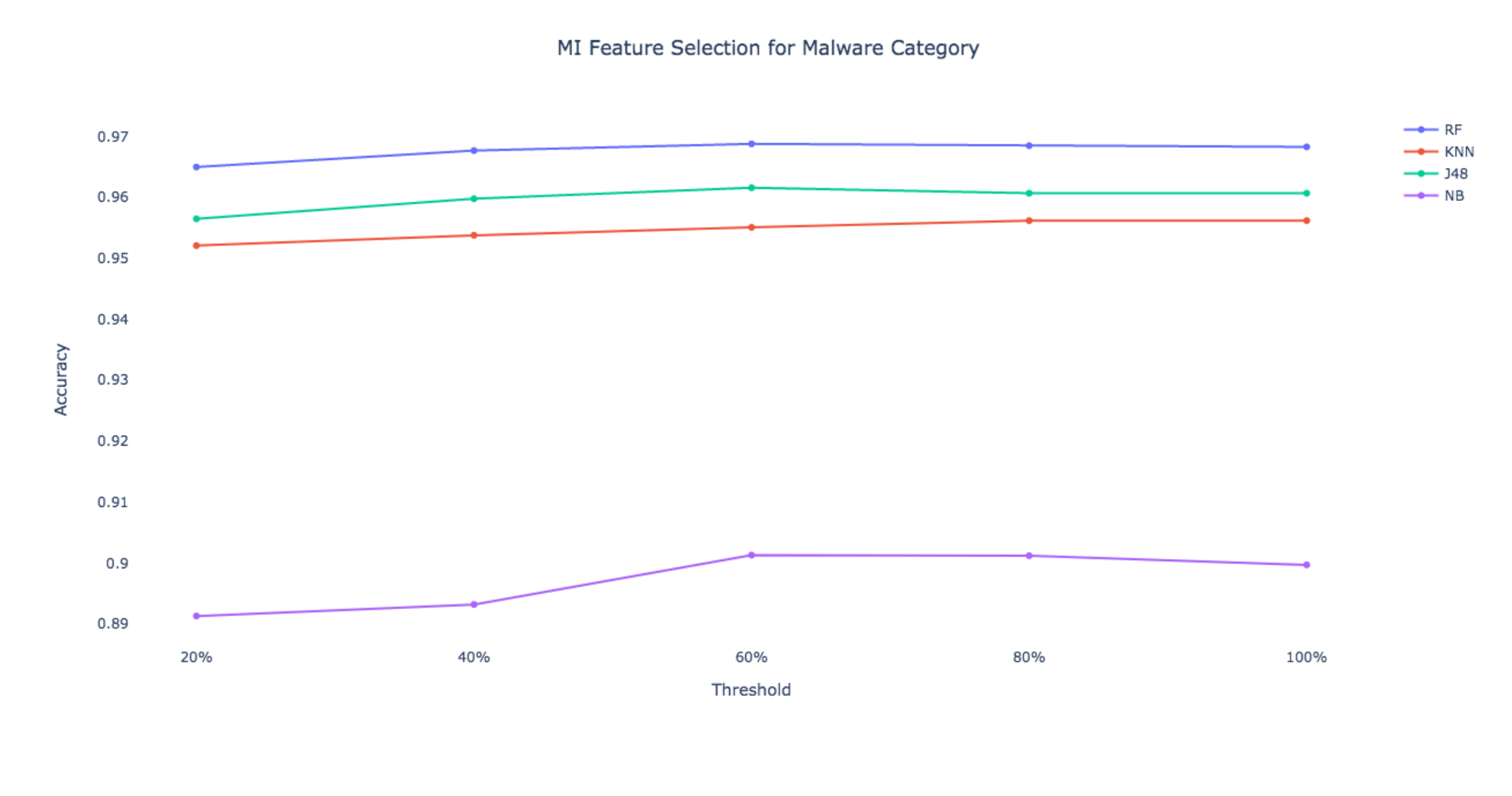

Figure 4: MI Feature Selection for Malware Category

Table 7: Comparison between accuracy and Threshold percentage for MI Feature Selection for Malware Category

| Threshold | Accuracy | | | |
|---|---|---|---|---|
| | RF | KNN | J48 | NB |
| 20% | 0.965 | 0.9521 | 0.9565 | 0.8913 |
| 40% | 0.9677 | 0.9538 | 0.9598 | 0.8932 |
| 60% | 0.9688 | 0.9551 | 0.9616 | 0.9013 |
| 80% | 0.9685 | 0.9562 | 0.9607 | 0.9012 |
| 100% | 0.9683 | 0.9562 | 0.9607 | 0.8997 |

2- After the first step: we can see that the best feature selection method is MI at threshold 60% (60% means we will use 60% of Features not all features which be 85 features not 141 features) as shown in Table 8.

Table 8: Selected Features of Android Malware Category Detection

| InfoGain | FeatureNumber | FeatureName |
|---|---|---|
| 0.279381731 | F76 | API_Crypto-Hash_java.security.MessageDigest_digest |
| 0.263039445 | F78 | API_DeviceInfo_android.telephony.TelephonyManager_getDeviceId |
| 0.243956604 | F97 | API_Network_java.net.URL_openConnection |
| 0.243153066 | F77 | API_Crypto-Hash_java.security.MessageDigest_update |
| 0.222883591 | F84 | API_DeviceInfo_android.net.wifi.WifiInfo_getMacAddress |
| 0.220628227 | F79 | API_DeviceInfo_android.telephony.TelephonyManager_getSubscriberId |
| 0.21562443 | F71 | API_Binder_android.app.ContextImpl_registerReceiver |
| 0.214375243 | F130 | Network_TotalReceivedPackets |
| 0.213290071 | F132 | Network_TotalTransmittedPackets |
| 0.21039932 | F70 | API_IPC_android.content.ContextWrapper_registerReceiver |
| 0.206074667 | F129 | Network_TotalReceivedBytes |
| 0.201332777 | F131 | Network_TotalTransmittedBytes |
| 0.19787337 | F103 | API_DexClassLoader_dalvik.system.BaseDexClassLoader_findLibrary |
| 0.194028256 | F72 | API_Binder_android.app.ActivityThread_handleReceiver |
| 0.180870419 | F27 | API_Command_java.lang.Runtime_exec |
| 0.180468905 | F28 | API_Command_java.lang.ProcessBuilder_start |
| 0.177479527 | F117 | API_DeviceData_android.content.ContentResolver_registerContentObserver |
| 0.173745799 | F109 | API_Base64_android.util.Base64_encodeToString |
| 0.170957536 | F108 | API_Base64_android.util.Base64_encode |
| 0.155062788 | F49 | API_Database_android.database.sqlite.SQLiteDatabase_execSQL |
| 0.15416418 | F99 | API_Network_com.android.okhttp.internal.huc.HttpURLConnectionImpl_getInputStream |
| 0.149961045 | F127 | API_DeviceData_android.app.ApplicationPackageManager_getInstalledPackages |
| 0.14943168 | F6 | Memory_PrivateClean |
| 0.147588947 | F60 | API_Database_android.database.sqlite.SQLiteDatabase_rawQueryWithFactory |
| 0.146474533 | F133 | Battery_wakelock |
| 0.144895957 | F59 | API_Database_android.database.sqlite.SQLiteDatabase_rawQuery |
| 0.144064408 | F113 | API_SystemManager_android.content.BroadcastReceiver_abortBroadcast |
| 0.142990688 | F51 | API_Database_android.database.sqlite.SQLiteDatabase_getPath |
| 0.142226391 | F45 | API_FileIO_android.content.ContextWrapper_deleteFile |
| 0.141621284 | F107 | API_Base64_android.util.Base64_decode |
| 0.141045198 | F3 | Memory_SharedDirty |
| 0.140401518 | F106 | API_DexClassLoader_dalvik.system.DexClassLoader_$init |
| 0.140157825 | F9 | Memory_HeapAlloc |
| 0.139557583 | F2 | Memory_PssClean |
| 0.13883735 | F74 | API_Crypto_javax.crypto.spec.SecretKeySpec_$init |
| 0.137096948 | F82 | API_DeviceInfo_android.telephony.TelephonyManager_getNetworkOperatorName |
| 0.135962761 | F128 | API__sessions |
| 0.132956747 | F8 | Memory_HeapSize |
| 0.13141196 | F134 | Battery_service |
| 0.13077421 | F68 | API_IPC_android.content.ContextWrapper_startService |
| 0.129308546 | F55 | API_Database_android.database.sqlite.SQLiteDatabase_openDatabase |
| 0.128014556 | F63 | API_Database_android.database.sqlite.SQLiteDatabase_compileStatement |
| 0.126900853 | F110 | API_SystemManager_android.app.ApplicationPackageManager_setComponentEnabledSetting |
| 0.126732667 | F1 | Memory_PssTotal |
| 0.12508465 | F5 | Memory_SharedClean |
| 0.125046105 | F65 | API_IPC_android.content.ContextWrapper_sendBroadcast |
| 0.12474779 | F4 | Memory_PrivateDirty |
| 0.121842371 | F75 | API_Crypto_javax.crypto.Cipher_doFinal |
| 0.120845917 | F12 | Memory_ViewRootImpl |
| 0.119304676 | F42 | API_FileIO_libcore.io.IoBridge_open |
| 0.118579678 | F52 | API_Database_android.database.sqlite.SQLiteDatabase_insert |
| 0.112506199 | F54 | API_Database_android.database.sqlite.SQLiteDatabase_insertWithOnConflict |
| 0.111304217 | F13 | Memory_AppContexts |
| 0.110725121 | F20 | Memory_ParcelCount |
| 0.109180263 | F126 | API_DeviceData_android.os.SystemProperties_get |
| 0.106588875 | F17 | Memory_LocalBinders |
| 0.105929247 | F101 | API_DexClassLoader_dalvik.system.BaseDexClassLoader_findResource |
| 0.104388236 | F15 | Memory_Assets |
| 0.103909499 | F114 | API_SMS_android.telephony.SmsManager_sendTextMessage |
| 0.099831594 | F61 | API_Database_android.database.sqlite.SQLiteDatabase_update |
| 0.09972831 | F62 | API_Database_android.database.sqlite.SQLiteDatabase_updateWithOnConflict |
| 0.099079597 | F31 | API_WebView_android.webkit.WebView_loadUrl |
| 0.09862924 | F90 | API_DeviceInfo_android.telephony.TelephonyManager_getNetworkCountryIso |
| 0.098499293 | F69 | API_IPC_android.content.ContextWrapper_stopService |
| 0.098098859 | F119 | API_DeviceData_android.content.ContentResolver_delete |
| 0.09782067 | F11 | Memory_Views |
| 0.097514566 | F14 | Memory_Activities |
| 0.097278565 | F73 | API_Binder_android.app.Activity_startActivity |
| 0.095838991 | F57 | API_Database_android.database.sqlite.SQLiteDatabase_query |
| 0.094695271 | F58 | API_Database_android.database.sqlite.SQLiteDatabase_queryWithFactory |
| 0.091703871 | F116 | API_DeviceData_android.content.ContentResolver_query |
| 0.089843238 | F19 | Memory_ParcelMemory |
| 0.088076596 | F21 | Memory_DeathRecipients |
| 0.087579289 | F18 | Memory_ProxyBinders |
| 0.086125649 | F118 | API_DeviceData_android.content.ContentResolver_insert |
| 0.085140684 | F23 | Memory_WebViews |
| 0.084946191 | F43 | API_FileIO_android.content.ContextWrapper_openFileInput |
| 0.081769714 | F89 | API_DeviceInfo_android.telephony.TelephonyManager_getSimSerialNumber |
| 0.080831523 | F10 | Memory_HeapFree |
| 0.076774514 | F80 | API_DeviceInfo_android.telephony.TelephonyManager_getLine1Number |
| 0.076042098 | F115 | API_SMS_android.telephony.SmsManager_sendMultipartTextMessage |
| 0.074412145 | F141 | Process_total |
| 0.069716101 | F34 | API_WebView_android.webkit.WebView_addJavascriptInterface |
| 0.064059781 | F67 | API_IPC_android.content.ContextWrapper_startActivity |
| 0.056480141 | F81 | API_DeviceInfo_android.telephony.TelephonyManager_getNetworkOperator |

3- Machine Learning Classifiers Results as shown in Table 9 and in Figure 5,6,7, and 8. Note, we run each experiment 10 time and get the average. Finally, we are note that the best Classifier is Random Forest (RF) in this case.

Table 9: ML Results for Malware Category Classification

| Classifier | Accuracy | Recall | FPR | FNR | Time (seconds) |
|---|---|---|---|---|---|
| RF | 0.9689 | 0.6646 | 0.0185 | 0.3354 | 987.16 |
| KNN | 0.9553 | 0.5563 | 0.0263 | 0.4437 | 1810.58 |
| J48 | 0.9616 | 0.6445 | 0.0222 | 0.3555 | 1389.67 |
| NB | 0.9015 | 0.3195 | 0.0512 | 0.6805 | 189.94 |
| MLP | 0.9632 | 0.5838 | 0.0217 | 0.4162 | 40874.48 |
| SVM | 0.9073 | 0.1202 | 0.0615 | 0.8798 | 23526.07 |
| AB | 0.9102 | 0.2897 | 0.0514 | 0.7103 | 3178.18 |
| LR | 0.9296 | 0.2848 | 0.0441 | 0.7152 | 1599.46 |

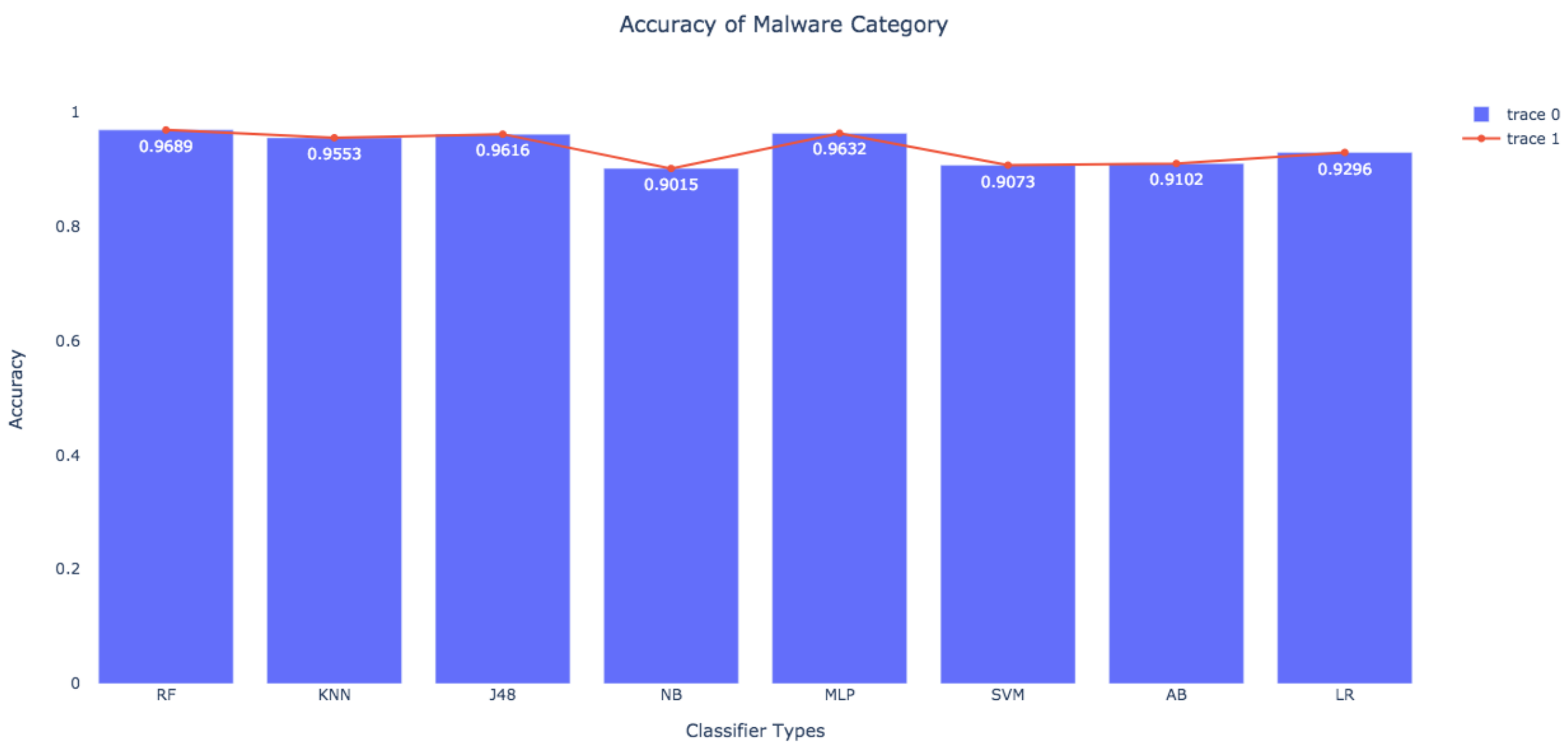

Figure 5: Accuracy of Android Malware Category

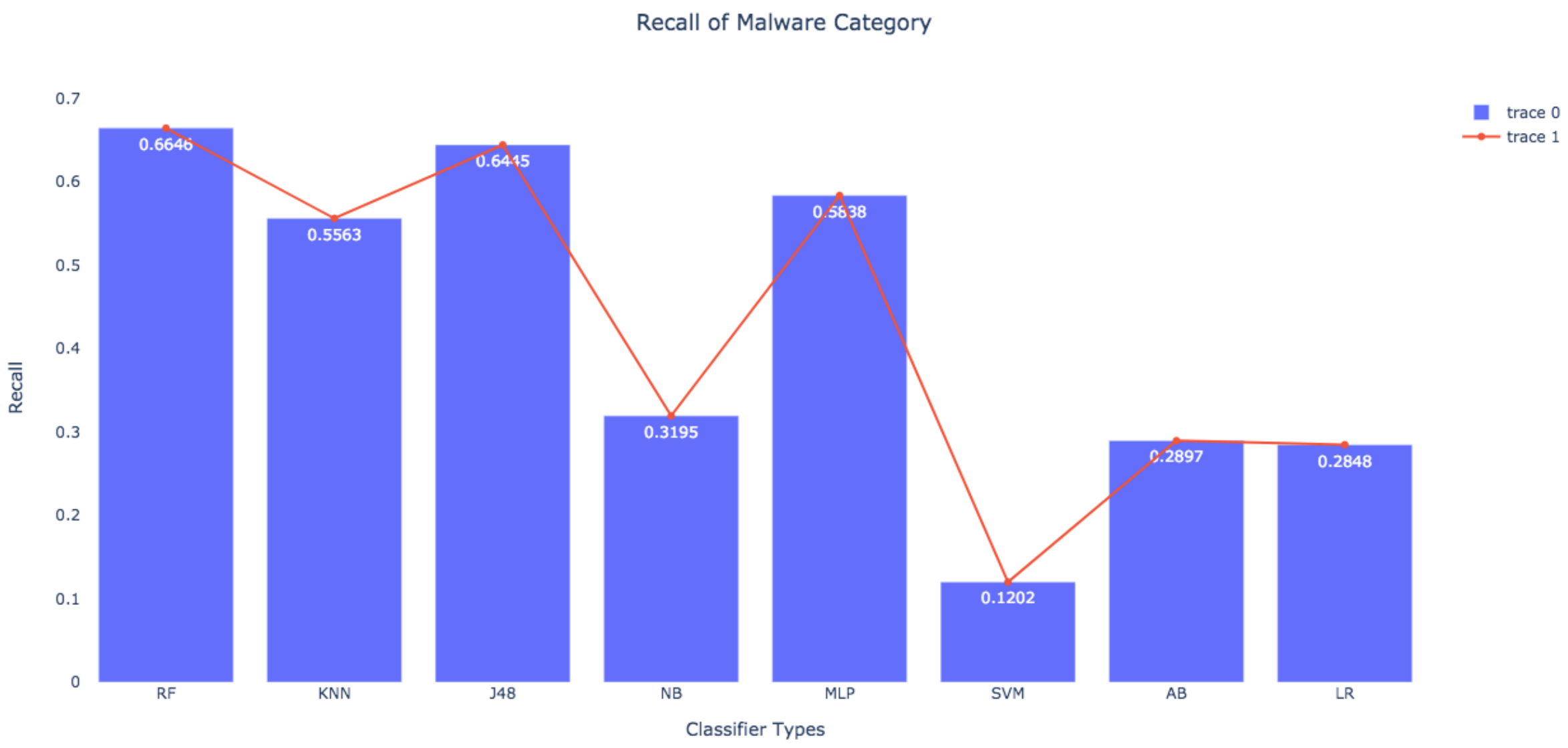

Figure 6: Recall of Android Malware Category

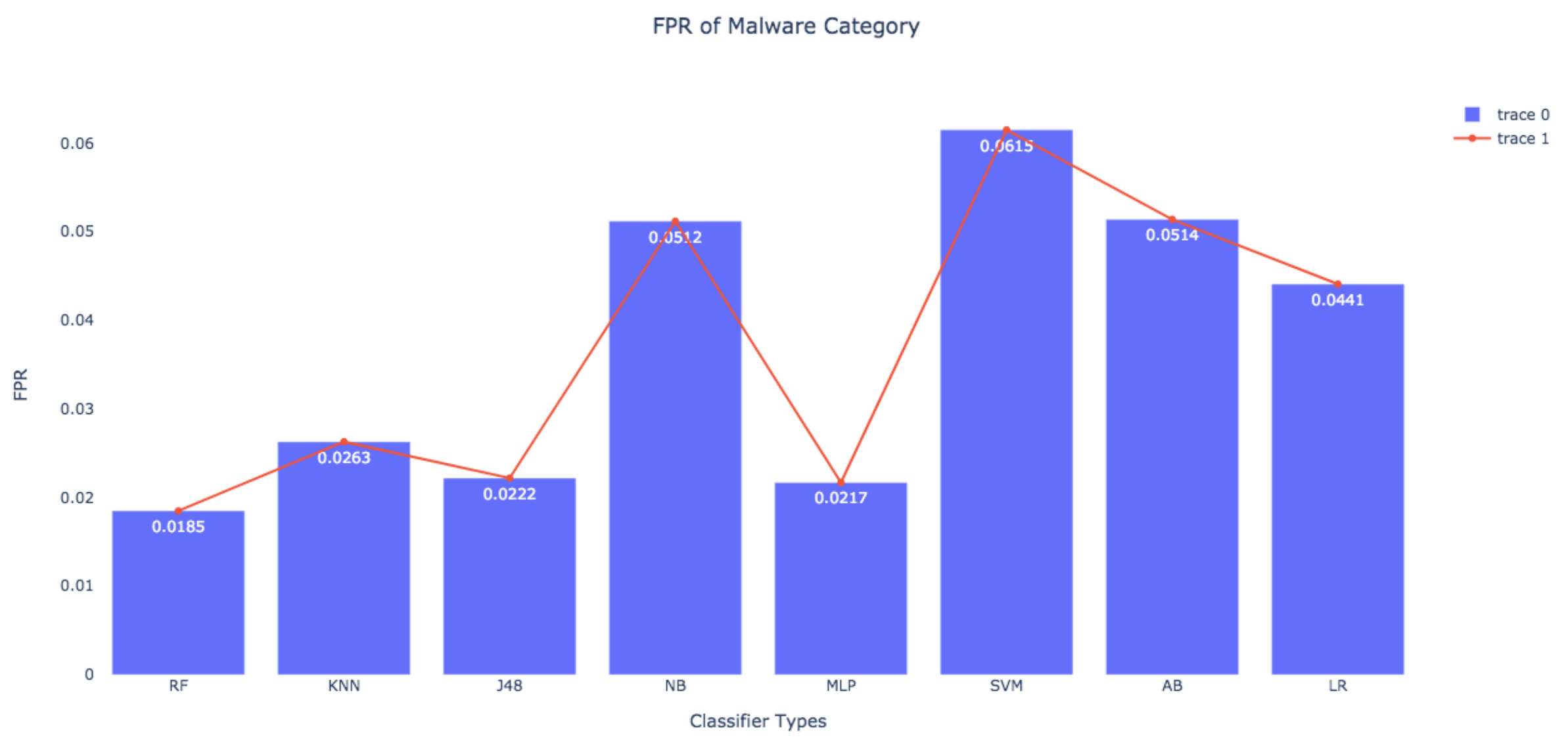

Figure 7: FPR of Android Malware Category

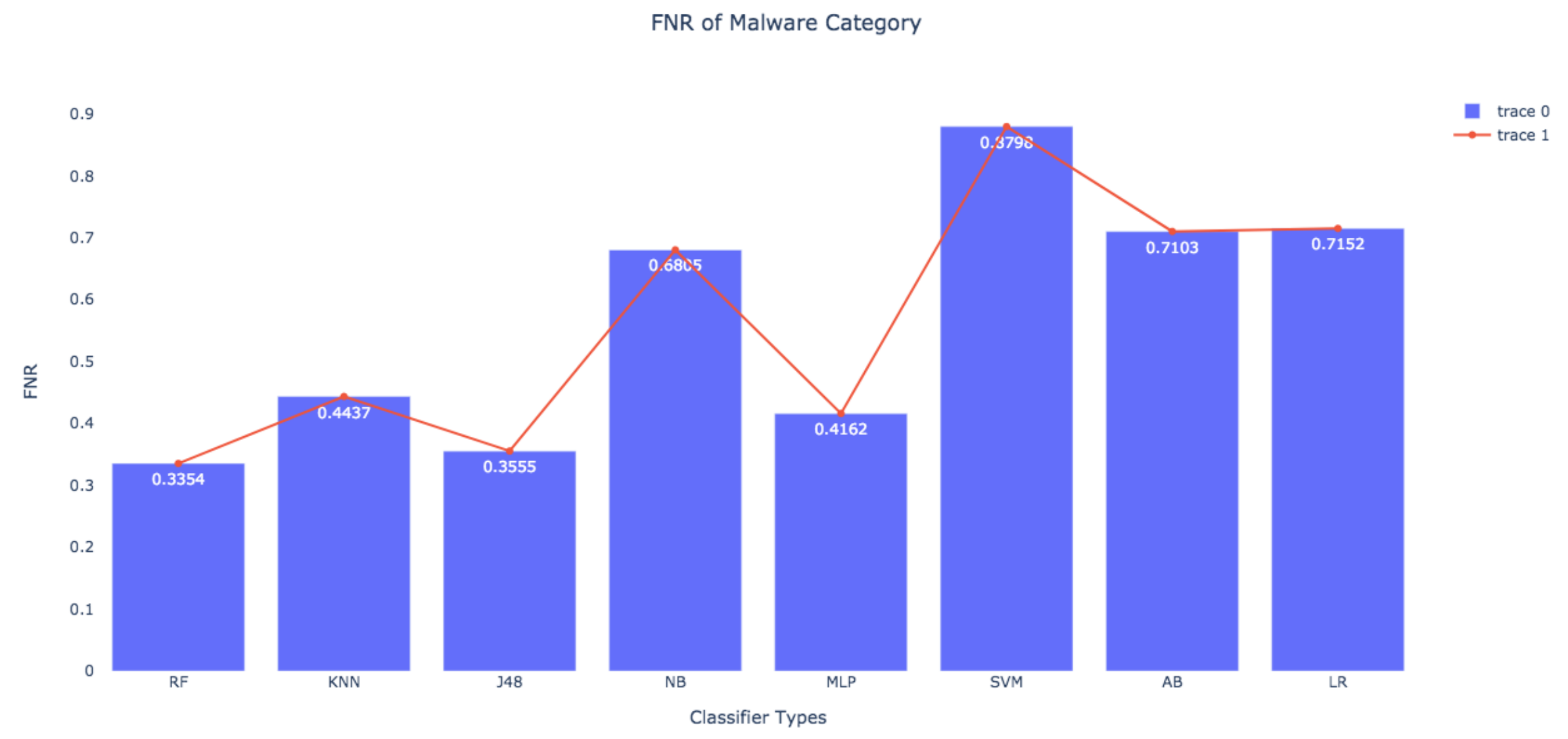

Figure 8: FNR of Android Malware Category

**Malware Family Detection**

1- Comparison between Feature selection methods Chi2 and MI as shown in Figure 9, 10 and Table 10, 11 to determine the best feature selection method and determine the best threshold.

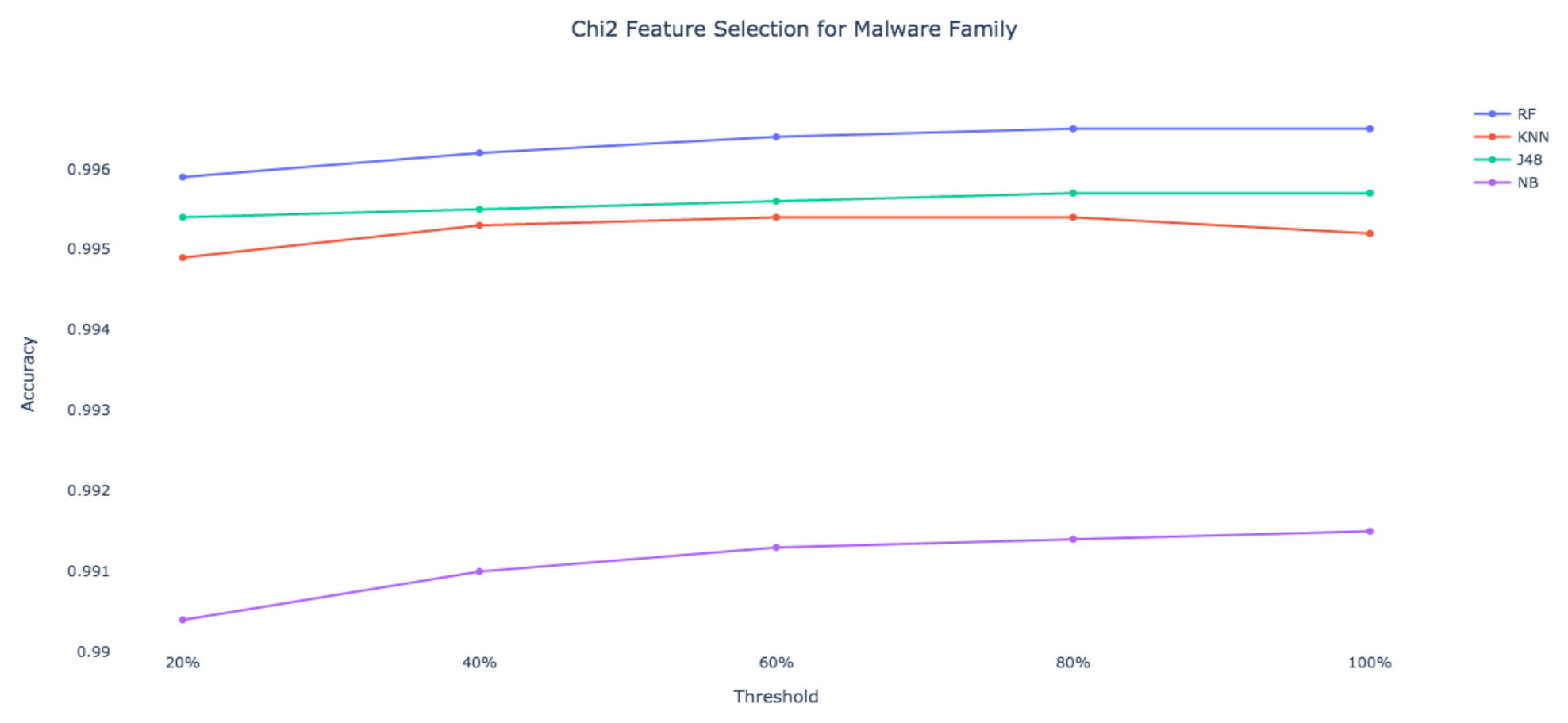

Figure 5: Chi2 Feature Selection for Malware Family

Table 10: Comparison between accuracy and Threshold percentage for Chi2 Feature Selection for Malware Family

| Threshold | Accuracy | | | |
|---|---|---|---|---|
| | RF | KNN | J48 | NB |
| 20% | 0.9959 | 0.9949 | 0.9954 | 0.9904 |
| 40% | 0.9962 | 0.9953 | 0.9955 | 0.991 |
| 60% | 0.9964 | 0.9954 | 0.9956 | 0.9913 |
| 80% | 0.9965 | 0.9954 | 0.9957 | 0.9914 |
| 100% | 0.9965 | 0.9952 | 0.9957 | 0.9915 |

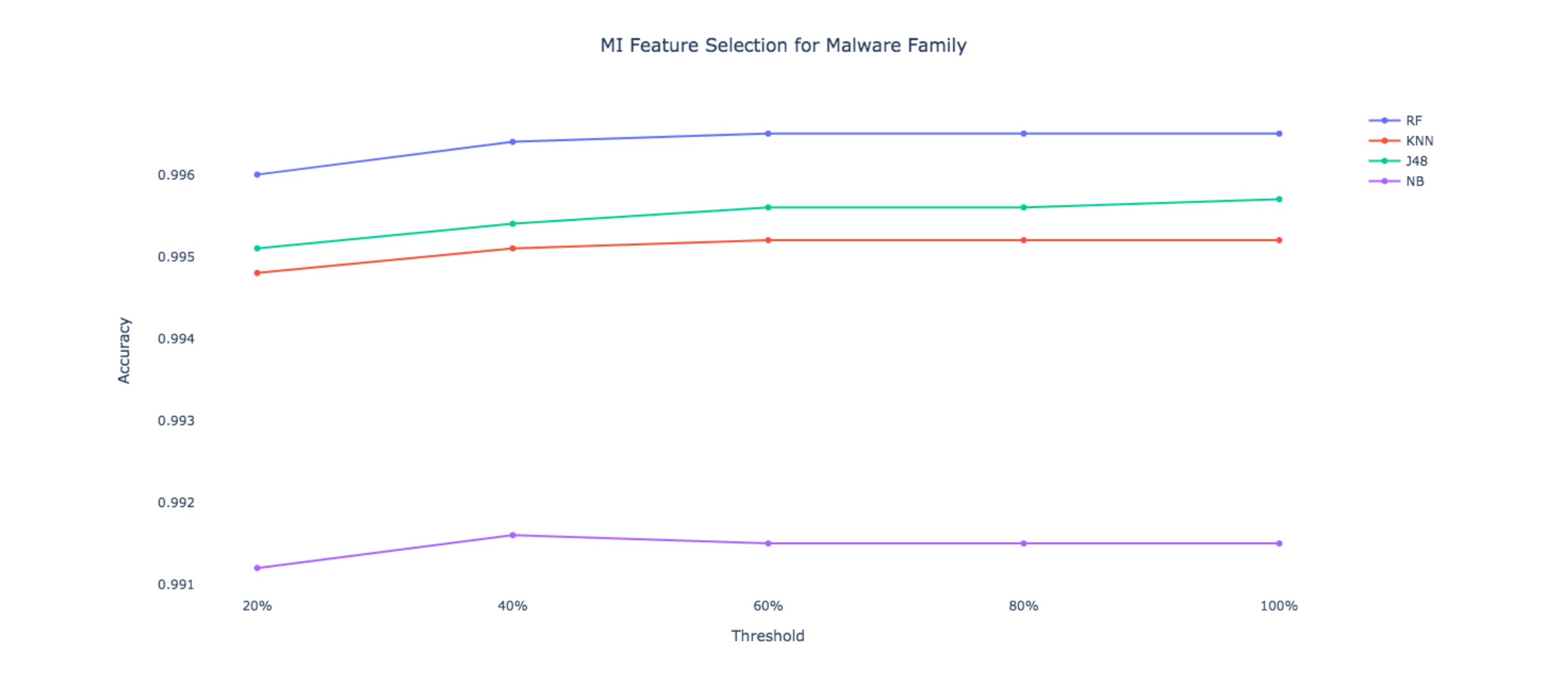

Figure 6: MI Feature Selection for Malware Family

Table 11: Comparison between accuracy and Threshold percentage for MI Feature Selection for Malware Family

| Threshold | Accuracy | | | |
|---|---|---|---|---|
| | RF | KNN | J48 | NB |
| 20% | 0.996 | 0.9948 | 0.9951 | 0.9912 |
| 40% | 0.9964 | 0.9951 | 0.9954 | 0.9916 |
| 60% | 0.9965 | 0.9952 | 0.9956 | 0.9915 |
| 80% | 0.9965 | 0.9952 | 0.9956 | 0.9915 |
| 100% | 0.9965 | 0.9952 | 0.9957 | 0.9915 |

2- After the first step: we can see that the best feature selection method is Chi2 at threshold 80% (80% means we will use 80% of Features not all features which be 113 features not 141 features) as shown in Table 12.

Table 12: Selected Features of Android Malware Family

| Chi2 | FeatureNumber | Feature Name |
|---|---|---|
| 57708.05015 | F69 | API_IPC_android.content.ContextWrapper_stopService |
| 50472.50633 | F119 | API_DeviceData_android.content.ContentResolver_delete |
| 49217.34761 | F105 | API_DexClassLoader_dalvik.system.DexFile_loadClass |
| 47297 | F30 | API_JavaNativeInterface_java.lang.Runtime_load |
| 40970.83697 | F106 | API_DexClassLoader_dalvik.system.DexClassLoader_$init |
| 40060.79831 | F127 | API_DeviceData_android.app.ApplicationPackageManager_getInstalledPackages |
| 39806.42233 | F92 | API_DeviceInfo_android.os.Debug_isDebuggerConnected |
| 39337.06531 | F61 | API_Database_android.database.sqlite.SQLiteDatabase_update |
| 39309.00524 | F62 | API_Database_android.database.sqlite.SQLiteDatabase_updateWithOnConflict |
| 38951.39112 | F104 | API_DexClassLoader_dalvik.system.DexFile_loadDex |
| 34166.25851 | F107 | API_Base64_android.util.Base64_decode |
| 33419.43364 | F118 | API_DeviceData_android.content.ContentResolver_insert |
| 32117.96106 | F120 | API_DeviceData_android.accounts.AccountManager_getAccountsByType |
| 31818.27323 | F91 | API_DeviceInfo_android.telephony.TelephonyManager_getDeviceSoftwareVersion |
| 29700.26853 | F52 | API_Database_android.database.sqlite.SQLiteDatabase_insert |
| 29554.35432 | F114 | API_SMS_android.telephony.SmsManager_sendTextMessage |
| 29051.22872 | F27 | API_Command_java.lang.Runtime_exec |
| 28791.34577 | F66 | API_IPC_android.content.ContextWrapper_sendStickyBroadcast |
| 28691.98367 | F28 | API_Command_java.lang.ProcessBuilder_start |
| 28507.92623 | F67 | API_IPC_android.content.ContextWrapper_startActivity |
| 27378.73322 | F116 | API_DeviceData_android.content.ContentResolver_query |
| 26801.79426 | F34 | API_WebView_android.webkit.WebView_addJavascriptInterface |
| 26770.41162 | F45 | API_FileIO_android.content.ContextWrapper_deleteFile |
| 26767.27932 | F54 | API_Database_android.database.sqlite.SQLiteDatabase_insertWithOnConflict |
| 26465.55928 | F44 | API_FileIO_android.content.ContextWrapper_openFileOutput |
| 26371.54006 | F89 | API_DeviceInfo_android.telephony.TelephonyManager_getSimSerialNumber |
| 26274.90152 | F84 | API_DeviceInfo_android.net.wifi.WifiInfo_getMacAddress |
| 26030.67482 | F31 | API_WebView_android.webkit.WebView_loadUrl |
| 25740.52566 | F110 | API_SystemManager_android.app.ApplicationPackageManager_setComponentEnabledSetting |
| 25665.61605 | F79 | API_DeviceInfo_android.telephony.TelephonyManager_getSubscriberId |
| 24388.40579 | F125 | API_DeviceData_android.media.MediaRecorder_start |
| 24307.34595 | F82 | API_DeviceInfo_android.telephony.TelephonyManager_getNetworkOperatorName |
| 24222.32568 | F88 | API_DeviceInfo_android.telephony.TelephonyManager_getSimCountryIso |
| 23965.88747 | F115 | API_SMS_android.telephony.SmsManager_sendMultipartTextMessage |
| 23357.24313 | F65 | API_IPC_android.content.ContextWrapper_sendBroadcast |
| 23113.49845 | F60 | API_Database_android.database.sqlite.SQLiteDatabase_rawQueryWithFactory |
| 23111.82794 | F49 | API_Database_android.database.sqlite.SQLiteDatabase_execSQL |
| 22572.63353 | F59 | API_Database_android.database.sqlite.SQLiteDatabase_rawQuery |
| 21806.82936 | F22 | Memory_OpenSSLSockets |
| 21529.57081 | F57 | API_Database_android.database.sqlite.SQLiteDatabase_query |
| 21500.33004 | F108 | API_Base64_android.util.Base64_encode |
| 21498.4177 | F58 | API_Database_android.database.sqlite.SQLiteDatabase_queryWithFactory |
| 21495.69101 | F51 | API_Database_android.database.sqlite.SQLiteDatabase_getPath |
| 21315.9672 | F78 | API_DeviceInfo_android.telephony.TelephonyManager_getDeviceId |
| 21141.05866 | F109 | API_Base64_android.util.Base64_encodeToString |
| 21105.83589 | F43 | API_FileIO_android.content.ContextWrapper_openFileInput |
| 20906.89187 | F103 | API_DexClassLoader_dalvik.system.BaseDexClassLoader_findLibrary |
| 20887.53121 | F102 | API_DexClassLoader_dalvik.system.BaseDexClassLoader_findResources |
| 20829.83125 | F99 | API_Network_com.android.okhttp.internal.huc.HttpURLConnectionImpl_getInputStream |
| 20579.51974 | F83 | API_DeviceInfo_android.telephony.TelephonyManager_getSimOperatorName |
| 20249.65293 | F46 | API_Database_android.content.ContextWrapper_openOrCreateDatabase |
| 19684.70118 | F63 | API_Database_android.database.sqlite.SQLiteDatabase_compileStatement |
| 19440.11924 | F68 | API_IPC_android.content.ContextWrapper_startService |
| 19352.11794 | F55 | API_Database_android.database.sqlite.SQLiteDatabase_openDatabase |
| 19347.72313 | F97 | API_Network_java.net.URL_openConnection |
| 18919.95403 | F25 | API_Process_android.app.ActivityManager_killBackgroundProcesses |
| 18688.28626 | F23 | Memory_WebViews |
| 18619.30654 | F48 | API_Database_android.content.ContextWrapper_deleteDatabase |
| 18619.30654 | F50 | API_Database_android.database.sqlite.SQLiteDatabase_deleteDatabase |
| 18437.31535 | F72 | API_Binder_android.app.ActivityThread_handleReceiver |
| 18421.01308 | F42 | API_FileIO_libcore.io.IoBridge_open |
| 18353.99334 | F101 | API_DexClassLoader_dalvik.system.BaseDexClassLoader_findResource |
| 18347.90232 | F121 | API_DeviceData_android.accounts.AccountManager_getAccounts |
| 18244.98341 | F81 | API_DeviceInfo_android.telephony.TelephonyManager_getNetworkOperator |
| 17540.4426 | F74 | API_Crypto_javax.crypto.spec.SecretKeySpec_$init |
| 17337.77626 | F90 | API_DeviceInfo_android.telephony.TelephonyManager_getNetworkCountryIso |
| 17283.27427 | F80 | API_DeviceInfo_android.telephony.TelephonyManager_getLine1Number |
| 17234.77648 | F75 | API_Crypto_javax.crypto.Cipher_doFinal |
| 17224.47623 | F77 | API_Crypto-Hash_java.security.MessageDigest_update |
| 16986.64655 | F33 | API_WebView_android.webkit.WebView_loadDataWithBaseURL |
| 16740.4292 | F73 | API_Binder_android.app.Activity_startActivity |
| 16611.61814 | F76 | API_Crypto-Hash_java.security.MessageDigest_digest |
| 16308.59134 | F117 | API_DeviceData_android.content.ContentResolver_registerContentObserver |
| 16235.66316 | F21 | Memory_DeathRecipients |
| 15668.41536 | F112 | API_SystemManager_android.telephony.TelephonyManager_listen |
| 15419.6639 | F71 | API_Binder_android.app.ContextImpl_registerReceiver |
| 15302.24083 | F85 | API_DeviceInfo_android.net.wifi.WifiInfo_getBSSID |
| 15258.04617 | F70 | API_IPC_android.content.ContextWrapper_registerReceiver |
| 15250.91298 | F134 | Battery_service |
| 15002.86453 | F111 | API_SystemManager_android.app.NotificationManager_notify |
| 14954.43259 | F113 | API_SystemManager_android.content.BroadcastReceiver_abortBroadcast |
| 14703.12012 | F128 | API__sessions |
| 14589.56087 | F35 | API_WebView_android.webkit.WebView_evaluateJavascript |
| 14400.46125 | F130 | Network_TotalReceivedPackets |
| 14321.25737 | F53 | API_Database_android.database.sqlite.SQLiteDatabase_insertOrThrow |
| 14276.58698 | F129 | Network_TotalReceivedBytes |
| 13739.00758 | F131 | Network_TotalTransmittedBytes |
| 13292.21682 | F132 | Network_TotalTransmittedPackets |
| 12066.08593 | F86 | API_DeviceInfo_android.net.wifi.WifiInfo_getIpAddress |
| 11441.57593 | F47 | API_Database_android.content.ContextWrapper_databaseList |
| 11421.7097 | F11 | Memory_Views |
| 11297.80636 | F8 | Memory_HeapSize |
| 11230.70778 | F20 | Memory_ParcelCount |
| 11076.80984 | F17 | Memory_LocalBinders |
| 10956.96574 | F56 | API_Database_android.database.sqlite.SQLiteDatabase_openOrCreateDatabase |
| 10869.82689 | F9 | Memory_HeapAlloc |
| 10828.40598 | F19 | Memory_ParcelMemory |
| 10569.45803 | F12 | Memory_ViewRootImpl |
| 10440.21363 | F126 | API_DeviceData_android.os.SystemProperties_get |
| 10342.44211 | F4 | Memory_PrivateDirty |
| 10049.13891 | F5 | Memory_SharedClean |
| 10008.52773 | F41 | API_WebView_android.webkit.WebView_setWebContentsDebuggingEnabled |
| 9826.357244 | F36 | API_WebView_android.webkit.WebView_postUrl |
| 9772.164141 | F32 | API_WebView_android.webkit.WebView_loadData |
| 9748.940512 | F6 | Memory_PrivateClean |
| 9542.278545 | F18 | Memory_ProxyBinders |
| 9354.718589 | F2 | Memory_PssClean |
| 9257.16105 | F1 | Memory_PssTotal |
| 8910.559328 | F10 | Memory_HeapFree |
| 8798.604277 | F13 | Memory_AppContexts |
| 8038.354522 | F14 | Memory_Activities |
| 7727.771459 | F3 | Memory_SharedDirty |
| 6098.046416 | F123 | API_DeviceData_android.location.Location_getLongitude |

3- Machine Learning Classifiers Results as shown in Table 13 and Figure 9,10,11, and 12. Note, we run each experiment 10 time and get the average. Finally, we are note also that the best Classifier is Random Forest (RF) in this case.

Table 13: ML Results for Malware Family Classification

| Classifier | Accuracy | Recall | FPR | FNR | Time (seconds) |
|---|---|---|---|---|---|
| RF | 0.9965 | 0.3627 | 0.0018 | 0.6373 | 312.39 |
| KNN | 0.9954 | 0.2697 | 0.0024 | 0.7303 | 922.07 |
| J48 | 0.9957 | 0.3421 | 0.0022 | 0.6579 | 470.99 |
| NB | 0.9914 | 0.2887 | 0.0043 | 0.7113 | 113.49 |
| MLP | 0.9960 | 0.3264 | 0.0021 | 0.6736 | 63165.61 |
| SVM | 0.9911 | 0.0262 | 0.0052 | 0.9738 | 23917.79 |
| AB | 0.9905 | 0.0187 | 0.0056 | 0.9813 | 6706.97 |
| LR | 0.9933 | 0.0844 | 0.0037 | 0.9156 | 6251.64 |

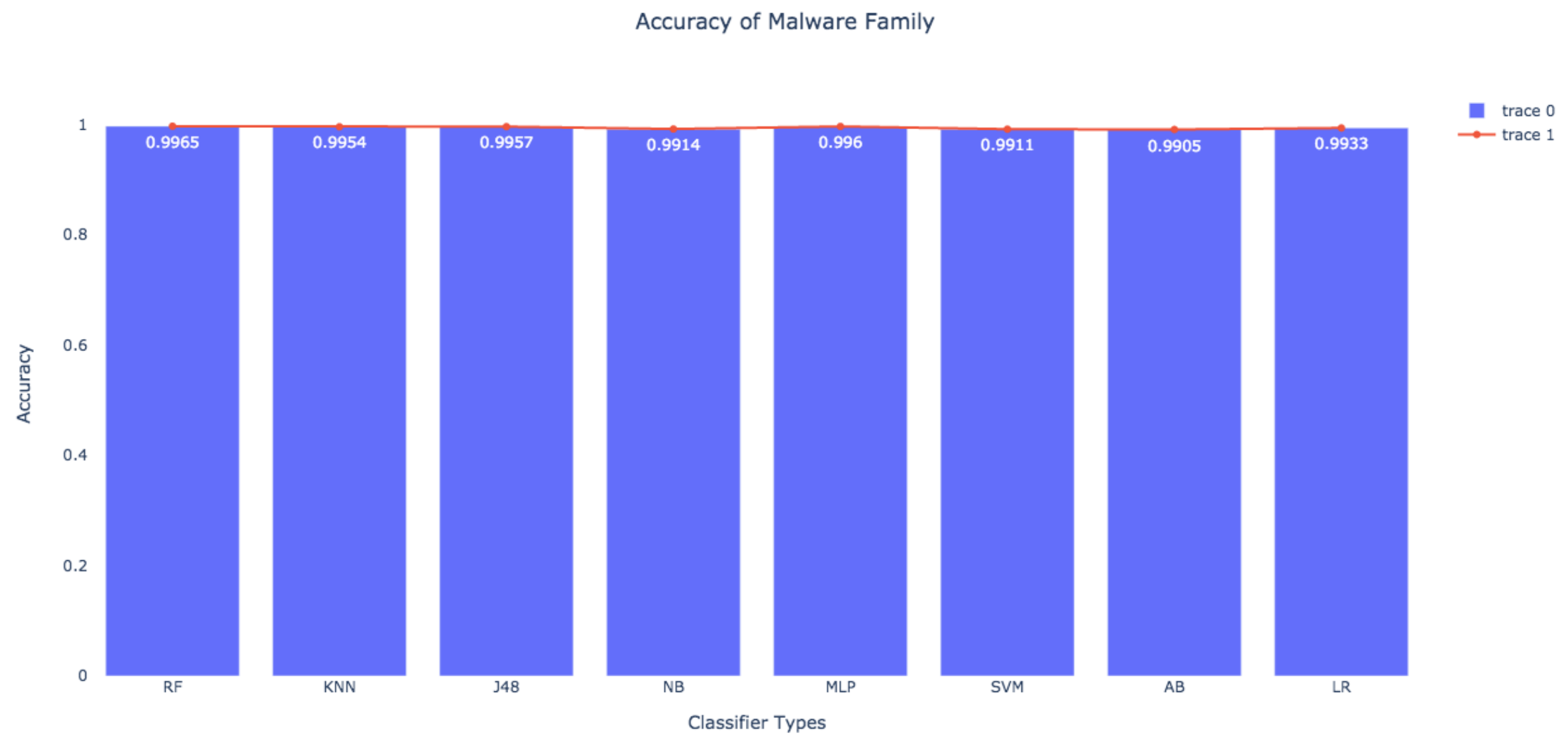

Figure 10: Accuracy of Android Family

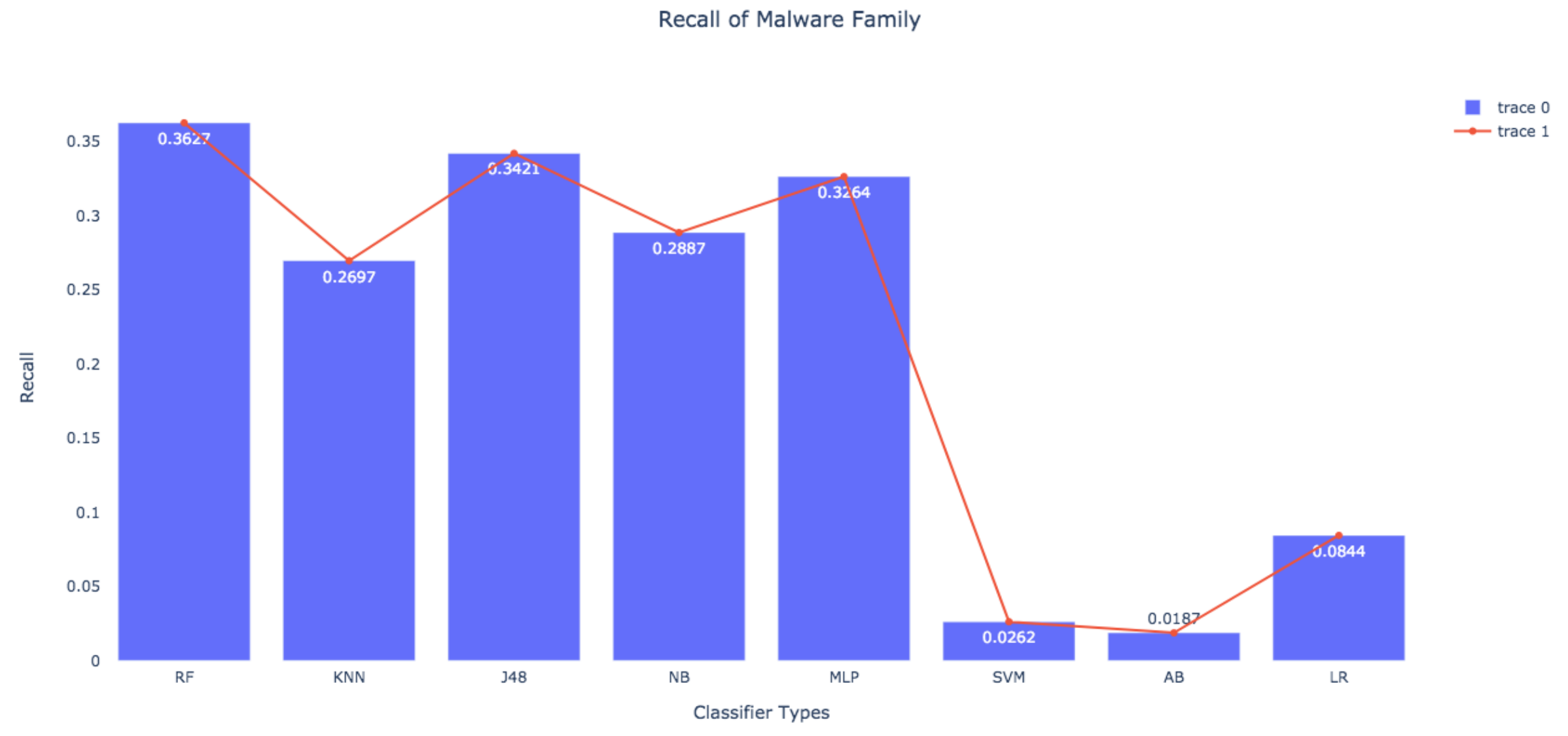

Figure 11: Recall of Android Malware Family

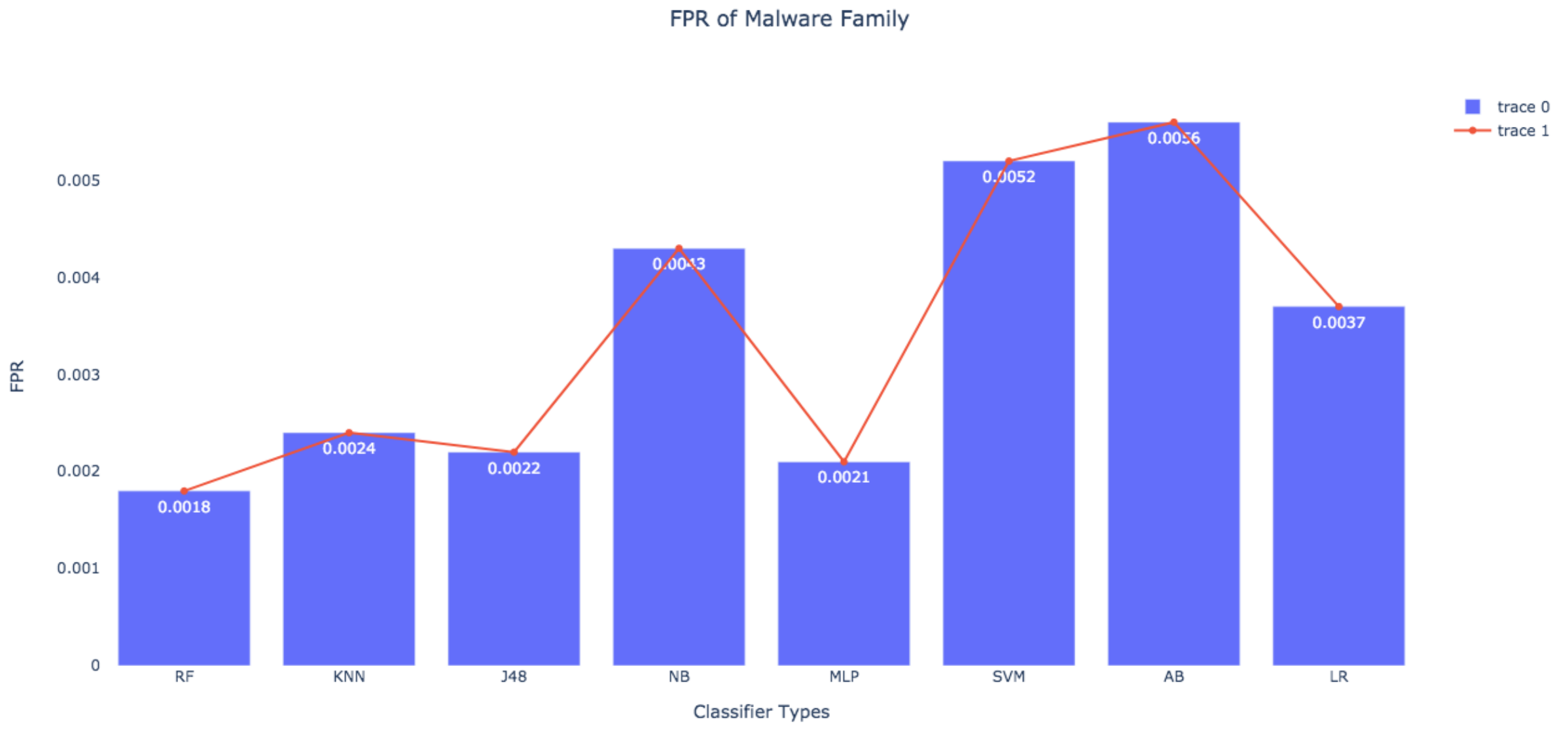

Figure 12: FPR of Android Malware Family

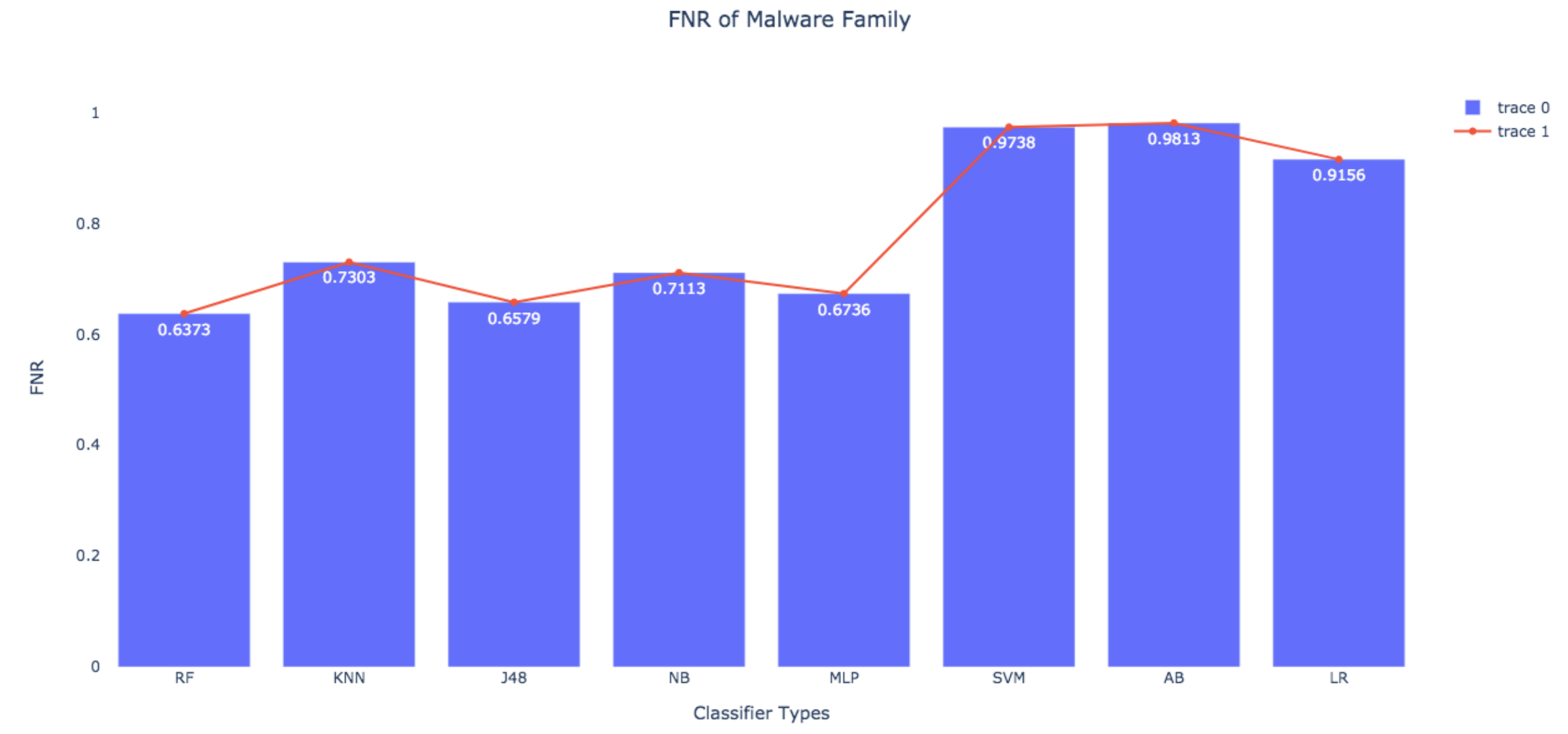

Figure 13: FNR of Android Malware Family

**Comparative Analysis**

We compare our experimental results with other works as in Table 14, and 15. We found that our proposed model is the best for many reason we use large dataset, latest version of dataset and we achieve high accuracy in case Malware category classification and Malware family classification.

Table 14: Result comparison of malware category classification

| Dataset | Accuracy |
|---|---|
| [31] | 83.3% (RF) |
| [44] | 49.5% (KNN) |
| [45] | 80.2% (RF) |
| [46] | 82.2% (Deep ANN) |
| Our Model | 96.89% (RF) |

Table 15: Result comparison of malware family classification

| Dataset | Accuracy |
|---|---|
| [31] | 59.7% (RF) |
| [44] | 27.24 % (KNN) |
| [46] | 65% (NB) |
| Our Model | 99.65% (RF) |

# Conclusion

Android malware is one of the most dangerous threats on the internet, and its prevalence has increased dramatically in recent years. Experts in cybersecurity face an open problem. There are a variety of machine learning-based approaches for detecting and classifying Android malware. This article offers a Machine Learning Model that uses feature selection and a Machine Learning Classifier to successfully perform malware classification and characterization techniques. Our model has shown encouraging results, with Malware Category Classification accuracy of over 96 % and Malware Family Classification accuracy of over 99 %. Furthermore, our algorithm accurately classifies the majority of the occurrences of all 180 malware families and 14 malware categories, demonstrating its efficacy. Because it can categorize a bigger dataset and malware families, our Machine Learning Model is scalable. It allows for multi-class characterization with high precision and a low rate of false negatives.

In the future, we plan to provide an online service that allows users to check if a program is malware or not before downloading it and determine its category and family. This measure would go a long way toward ensuring the security of an Android smartphone.

# References:

[1] Share of global smartphone shipments by operating system from 2014 to 2023: https://www.statista.com/statistics/272307/market-share-forecast-for-smartphone-operating-systems/

[2] Operating System Market Share Worldwide: https://gs.statcounter.com/os-market-share

[3] Cyber attacks on Android devices on the rise: https://www.gdatasoftware.com/blog/2018/11/31255-cyber-attacks-on-android-devices-on-the-rise

[4] I. Ideses, A. Neuberger, Adware detection and privacy control in mobile devices, in 2014 IEEE 28th Convention of Electrical & Electronics Engineers in Israel (IEEEI), IEEE, 2014, pp. 1–5.

[5] J.-S. Ko, J.-S. Jo, D.-H. Kim, S.-K. Choi, J. Kwak, Real time android ransomware detection by analyzed android applications, in: 2019 International Conference on Electronics, Information, and Communication (ICEIC), IEEE, 2019, pp. 1–5.

[6] M. Sikorski, A. Honig, Practical Malware Analysis: The Hands-On Guide to Dissecting Malicious Software, no starch press, 2012.

[7] F. Faghihi, M. Abadi, A. Tajoddin, Smsbothunter: A novel anomaly detection technique to detect sms botnets, in: 2018 15th International ISC (Iranian Society of Cryptology) Conference on Information Security and Cryptology (ISCISC), IEEE, 2018, pp. 1–6.

[8] Zhuo Ma, Haoran Ge, Yang Liu, Meng Zhao, and Jianfeng Ma. 2019. A Combination Method for Android Malware Detection Based on Control Flow Graphs and Machine Learning Algorithms. IEEE Access 7 (2019), 21235–21245. https://doi.org/10.1109/ACCESS.2019.2896003
[9] Recep Sinan Arslan, İbrahim Alper Doğru, and Necaattin Barişçi . 2019. Permission Based Malware Detection System for Android Using Machine Learning Techniques. International Journal of Software Engineering and Knowledge Engineering 29,1(2019),43 61. https://doi.org/10.1142/S0218194019500037
[10] F. Noorbehbahani, F. Rasouli, M. Saberi, Analysis of machine learning techniques for ransomware detection, in: 2019 16th International ISC (Iranian Society of Cryptology) Conference on Information Security and Cryptology (ISCISC), IEEE, 2019, pp. 128–133.
[11] Android malware dataset (cicandmal2017 - first part), 2020, https://www. unb.ca/cic/datasets/andmal2017.html
[12] R. Chen, Y. Li, W. Fang, Android malware identification based on traffic analysis, in: International Conference on Artificial Intelligence and Security, Springer, 2019, pp. 293–303.
[13] Intrusion detection evaluation dataset (cicids2017), 2020, https://www. unb.ca/cic/datasets/ids-2017.html (Accessed: 2020-03-12).
[14] I. Sharafaldin, A.H. Lashkari, A.A. Ghorbani, Toward generating a new intrusion detection dataset and intrusion traffic characterization, in: ICISSP, 2018, pp. 108–116.
[15] M. Samara, E.-S.M. El-Alfy, Benchmarking open-source android mal- ware detection tools, in: 2019 2nd IEEE Middle East and North Africa COMMunications Conference (MENACOMM), IEEE, 2019, pp. 1–6.
[16] D. Maiorca, F. Mercaldo, G. Giacinto, C.A. Visaggio, F. Martinelli, R- PackDroid: API package-based characterization and detection of mobile ransomware, in: Proceedings of the Symposium on Applied Computing, 2017, pp. 1718–1723.
[17] Songhao Lou, Shaoyin Cheng, Jingjing Huang, and Fan Jiang. 2019. TFDroid: Android Malware Detection by Topics and Sensitive Data Flows Using Machine Learning Techniques. IEEE 2nd International Conference on Information and Computer Technologies, Kahului, HI, USA(2019),30–36. https://doi.org/10.1109/INFOCT.2019.8711179
[18] Suman R. Tiwari and Ravi U. Shukla. 2018. An Android Malware Detection Technique Based on Optimized Permissions and API. International Conference on Inventive Research in ComputingApplications(2018),2611–2616. https://doi.org/ 10.1109/ICCONS.2018.8662939
[19] Hanqing Zhang, Senlin Luo, Yifei Zhang, and Limin Pan. 2019. An Efficient Android Malware Detection System Based on Method Level Behavioral Semantic Analysis. IEEEAccess7(2019),69246–69256. https://doi.org/10.1109/ACCESS. 2019.2919796
[20] Enrico Mariconti, Lucky Onwuzurike, Panagiotis Andriotis, Emiliano De Cristofaro, Gordon Ross, and Gianluca Stringhini. 2017. MAMADROID: Detecting Android Malware by Building Markov Chains of Behavioral Models. (2017). https://arxiv.org/abs/1612.04433
[21] Suleiman Y. Yerima and Sarmadullah Khan. 2019. Longitudinal performance analysis of machine learning based Android malware detectors. International Conference on Cyber Security and Protection of Digital Services, Oxford, United Kingdom (2019),1–8. https://doi.org/10.1109/CyberSecPODS.2019.8885384
[22] Ming Fan, Xiapu Luo, Jun Liu, Meng Wang, Chunyin Nong, Qinghua Zheng, and Ting Liu. 2019. Graph Embedding based Familial Analysis of Android Malware using Unsupervised Learning. Proceedings of the 41st International Conference on Software Engineering(2019),771–782. https://doi.org/10.1109/ICSE.2019.00085
[23] Ali Feizollah, Nor Badrul Anuar, Rosli Salleh, Guillermo Suarez-Tangil, and Steven Furnell. 2017. AndroDialysis: Analysis of Android Intent Effectiveness in Malware Detection. Computers & Security 65 (2017), 121–134. https: //doi.org/10.1109/ICSE.2019.00085
[24] Wei Wang, Yuanyuan Li, Xing Wang, Jiqiang Liu, and Xiangliang Zhang. 2018. Detecting Android malicious apps and categorizing benign apps with ensemble of classifiers. Future

Generation Computer Systems 78 (2018), 987–994.
https://doi.org/10.1109/ICSE.2019.00085
[25] Mehmet Ali Atici, Seref Sagiroglu, and Ibrahim Alper Dogru. 2016. Android malware analysis approach based on control flow graphs and machine learning algorithms. 4th International Symposium on Digital Forensic and Security, LittleRock,AR(2016),26–31. https://doi.org/10.1109/ISDFS.2016.7473512
[26] KeXu,YingjiuLi,andRobertH.Deng.2016.ICCDetector:ICC-Based Malware Detectionon Android. IEEE Transactions on Information Forensics and Security11,6(2016),1252–1264. https://doi.org/10.1109/TIFS.2016.2523912
[27] Joshua Garcia, Mahmoud Hammad, and Sam Malek. 2016. Lightweight, Obfuscation-Resilient Detection and Family Identification of Android Malware. ACM Transactiosn on Software Engineering Methodology 26, 3 (2016).
[28] Heng Li, ShiYao Zhou, Wei Yuan, and Jiahuan Liand Henry Leung. 2019. Adversarial-Example Attacks Toward Android Malware Detection System. IEEE Systems Journal 14,1(2019),653–656. https://doi.org/10.1109/JSYST.2019.2906120
[29] Canadian Centre for Cyber Security. https://cyber.gc.ca/en/
[30] Francois Gagnon and Frederic Massicotte. 2017. Revisiting Static Analysis of Android Malware. Proceedings of the 10th USENIX Conference on Cyber Security Experimentation and Test, Vancouver, BC, Canada (2017).
[31] L. Taheri, A.F.A. Kadir, A.H. Lashkari, Extensible android malware detection and family classification using network-flows and api-calls, in: 2019 International Carnahan Conference on Security Technology (ICCST), IEEE, 2019, pp. 1–8.
[32] Understanding Android Malware Families (UAMF) – The Foundations (Article 1) https://www.itworldcanada.com/blog/understanding-android-malware-families-uamf-the-foundations-article-1/441562
[33] M.K.A. Abuthawabeh, K.W. Mahmoud, Android malware detection and categorization based on conversation-level network traffic features, in: 2019 International Arab Conference on Information Technology (ACIT), IEEE, 2019, pp. 42–47.
[34] A. Rehman Javed, Z. Jalil, S. Atif Moqurrab, S. Abbas, X. Liu, Ensemble adaboost classifier for accurate and fast detection of botnet attacks in connected vehicles, Trans. Emerg. Telecommun. Technol. (2020) e4088.
[35] J.-S. Ko, J.-S. Jo, D.-H. Kim, S.-K. Choi, J. Kwak, Real time android ransomware detection by analyzed android applications, in: 2019 Interna- tional Conference on Electronics, Information, and Communication (ICEIC), IEEE, 2019, pp. 1–5.
[36] N.T. Cam, V.-H. Pham, T. Nguyen, Detecting sensitive data leakage via inter- applications on android using a hybrid analysis technique, Cluster Comput. 22 (1) (2019) 1055–1064.
[37] Z. Yuan, Y. Lu, Y. Xue, Droiddetector: android malware characterization and detection using deep learning, Tsinghua Sci. Technol. 21 (1) (2016) 114–123.
[38] I. Ideses, A. Neuberger, Adware detection and privacy control in mobile devices, in: 2014 IEEE 28th Convention of Electrical & Electronics Engineers in Israel (IEEEI), IEEE, 2014, pp. 1–5.
[39] K. Sharma, B. Gupta, Towards privacy risk analysis in android applications using machine learning approaches, Int. J. E-Serv. Mob. Appl. (IJESMA) 11 (2) (2019) 1–21.
[40] F. Faghihi, M. Abadi, A. Tajoddin, Smsbothunter: A novel anomaly detection technique to detect sms botnets, in: 2018 15th International ISC (Iranian Society of Cryptology) Conference on Information Security and Cryptology (ISCISC), IEEE, 2018, pp. 1–6.

[41] F. Tchakounté, A.D. Wandala, Y. Tiguiane, Detection of android malware based on sequence alignment of permissions, Int. J. Comput. (IJC) 35 (1) (2019) 26–36.
[42] M. Sikorski, A. Honig, Practical Malware Analysis: The Hands-On Guide to Dissecting Malicious Software, no starch press, 2012.
[43] Y. Nishimoto, N. Kajiwara, S. Matsumoto, Y. Hori, K. Sakurai, Detection of android api call using logging mechanism within android framework, in: International Conference on Security and Privacy in Communication Systems, Springer, 2013, pp. 393–404.
[44] A.H. Lashkari, A.F.A. Kadir, L. Taheri, A.A. Ghorbani, Toward developing a systematic approach to generate benchmark android malware datasets and classification, in: 2018 International Carnahan Conference on Security Technology (ICCST), IEEE, 2018, pp. 1–7.
[45] M.K.A. Abuthawabeh, K.W. Mahmoud, Android malware detection and categorization based on conversation-level network traffic features, in: 2019 International Arab Conference on Information Technology (ACIT), IEEE, 2019, pp. 42–47.
[46] Syed Ibrahim Imtiaz, Saif ur Rehman, Abdul Rehman Javed, Zunera Jalil, Xuan Liu, Waleed S. Alnumay, DeepAMD: Detection and identification of Android malware using high-efficient Deep Artificial Neural Network, Future Generation Computer Systems, Volume 115, 2021, Pages 844-856, ISSN 0167-739X